\documentclass[final,1p,times]{elsarticle}

\usepackage[utf8]{inputenc}
\usepackage{amsmath,amssymb,mathtools}
\usepackage{tikz}
\usepackage{algorithm}
\usepackage{algpseudocode}
\usepackage{etoolbox} 
\usepackage{wrapfig}
\usepackage[color,all]{xy}
\usepackage{fancybox}
\usepackage{tabularx}
\usepackage{xargs}
\usepackage{hyperref}
\usepackage{subcaption}
\usepackage{fancyhdr}

\newcommand{\graphname}[3][]{\ensuremath{#2_{#3}^{#1}}}
\newcommand{\chain}[3]{(\overline{#1},{#2},{#3})}
\newcommandx{\cat}[2][2]{\mbox{\boldmath\ensuremath{\mathsf{#1}}\unboldmath}#2}
\newcommand{\chainname}[1]{\ensuremath{\mathcal{#1}}}
\newcommand{\source}[1][]{\mathit{sc}^{#1}}
\newcommand{\target}[1][]{\mathit{tg}^{#1}}
\newcommand{\elementname}[1]{\texttt{#1}}
\newcommand{\potencyseparator}{@}
\newcommand{\typename}[2]{\texttt{#1\potencyseparator#2}}
\newcommand{\typeof}[2][]{\ensuremath{\mathit{ty}^{#1}(#2)}}
\newcommand{\typegraph}[1]{\ensuremath{\mathit{TG}_{#1}}}
\newcommand{\typegraphof}[2][]{\ensuremath{\mathit{TG}^{#1}(#2)}}
\newcommand{\level}[1]{#1}

\newcommand{\typemorph}[3][]{\ifstrempty{#2}{\ensuremath{\tau^{#1}}}{\ensuremath{\tau_{#2,#3}^{#1}}}}
\newcommand{\domain}[1]{D(#1)}
\newcommand{\tc}[1]{\overline{#1}}
\newcommand{\partialmap}{{\mbox{\(\hspace{0.7em}\circ\hspace{-1.3em}\longrightarrow\hspace{0.1em}\)}}}
\newcommand{\difference}[2][]{\mathit{df}^{#1}(#2)}
\newcommand{\step}[1]{s_{#1}}

\newcommand{\algstyle}[1]{\ensuremath{\mathit{#1}}}
\newcommand{\chainmorph}[2][]{\ensuremath{\sigma_{#2}^{#1}}}

\newdefinition{definition}{Definition}
\newdefinition{theorem}{Theorem}
\newdefinition{remark}{Remark}
\newdefinition{lemma}{Lemma}
\newdefinition{corollary}{Corollary}
\newdefinition{proof}{Proof}

\usetikzlibrary{arrows,positioning,decorations.markings}

\journal{Journal of Logical and Algebraic Methods in Programming}

\begin{document}

\begin{frontmatter}

\title{Multilevel Coupled Model Transformations for Precise and Reusable Definition of Model Behaviour}

\author[HVL]{Fernando Macías}
\author[UIB]{Uwe Wolter}
\author[HVL]{Adrian Rutle}
\author[UMA]{Francisco Durán}
\author[UEX]{Roberto Rodriguez-Echeverria}
\address[HVL]{Western Norway University of Applied Sciences, Bergen, Norway}
\address[UIB]{University of Bergen, Bergen, Norway}
\address[UMA]{University of Málaga, M\'alaga, Spain}
\address[UEX]{University of Extremadura, Cáceres, Spain}

\begin{abstract}

The use of Domain-Specific Languages (DSLs) is a promising field for the development of tools tailored to specific problem spaces, effectively diminishing the complexity of hand-made software.
With the goal of making models as precise, simple and reusable as possible, we augment DSLs with concepts from multilevel modelling, where the number of abstraction levels are not limited.
This is particularly useful for DSL definitions with behaviour, whose concepts inherently belong to different levels of abstraction.
Here, models can represent the state of the modelled system and evolve using model transformations.
These transformations can benefit from a multilevel setting, becoming a precise and reusable definition of the semantics for behavioural modelling languages.
We present in this paper the concept of Multilevel Coupled Model Transformations, together with examples, formal definitions and tools to assess their conceptual soundness and practical value.

\end{abstract}


\end{frontmatter}

\fancyhf{}
\renewcommand{\headrulewidth}{0pt}
\renewcommand{\footrulewidth}{1pt}
\fancyfoot[L]{
\begin{wrapfigure}[2]{r}[0mm]{.28\columnwidth}
\vspace{-5mm}
\includegraphics[width=.28\columnwidth]{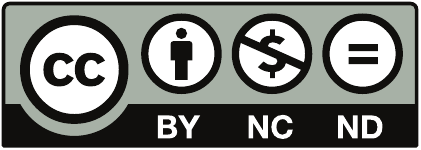}
\end{wrapfigure}
\textcopyright~2019. This manuscript version is made available
\newline
under the CC-BY-NC-ND 4.0 license \url{http://creativecommons.org/licenses/by-nc-nd/4.0/}
}%
\pagestyle{empty}
\thispagestyle{fancy}%

\section{Introduction}
\label{sec:introduction}

Model-driven software engineering (MDSE) is one of the emergent responses from the scientific and industrial communities to tackle the increasing complexity of software systems.
MDSE utilises abstractions for modelling different aspects of software systems, and treats models as first-class entities in all phases of software development.
There are quite a few studies which support gains in quality, performance, effectiveness, etc. as a result of using MDSE (see, e.g.,~\cite{WhittleHRBH13,WhittleHR14,BurdenHW14,MohagheghiGSFNF13,MohagheghiGSF13}).
However, using modelling to understand a domain, making the right abstractions, and including all the stakeholders in the development process, are without doubt the main gains of MDSE~\cite{BurdenHW14,MohagheghiFMFG08,Tolvanen016}.

According to empiric evaluations related to the status and practice of MDSE, the state-of-the-art modelling techniques and tools do a poor job in supporting software development activities~\cite{WhittleHR14}.
There is basically no consensus on modelling languages or tools --- MDSE-developers have reported more than 40 modelling languages and 100 tools as ``regularly used''~\cite{HutchinsonWRK11}.
Moreover, a recent study~\cite{Petre14,Petre13} found that software designers either did not use the Unified Modelling Language (UML)~\cite{uml} at all or used it only selectively and informally.
The main issue here is that most modelling approaches are developed without an appreciation for how people and organisations work~\cite{WhittleHR14}.

One way to increase the adoption of MDSE in practise is to develop modelling approaches which reflect the way software architects, developers and designers, as well as organisations, domain experts and stakeholders, handle abstraction and problem-solving.
We believe that domain-specific (meta)modelling (DSMM)~\cite{deLara2015,MohagheghiH10,Tolvanen016,Kelly08DSMBook} is an approach that could unite software modelling and abstraction, software design and architecture, and organisational studies.
This would help in filling the gap between these fields which ``could solve all kinds of problems and make modelling even more widely applicable than it currently is''~\cite{WhittleHR14}.

\begin{wrapfigure}[17]{r}[0mm]{.45\columnwidth}
	\centering
	\includegraphics[width=0.45\textwidth]{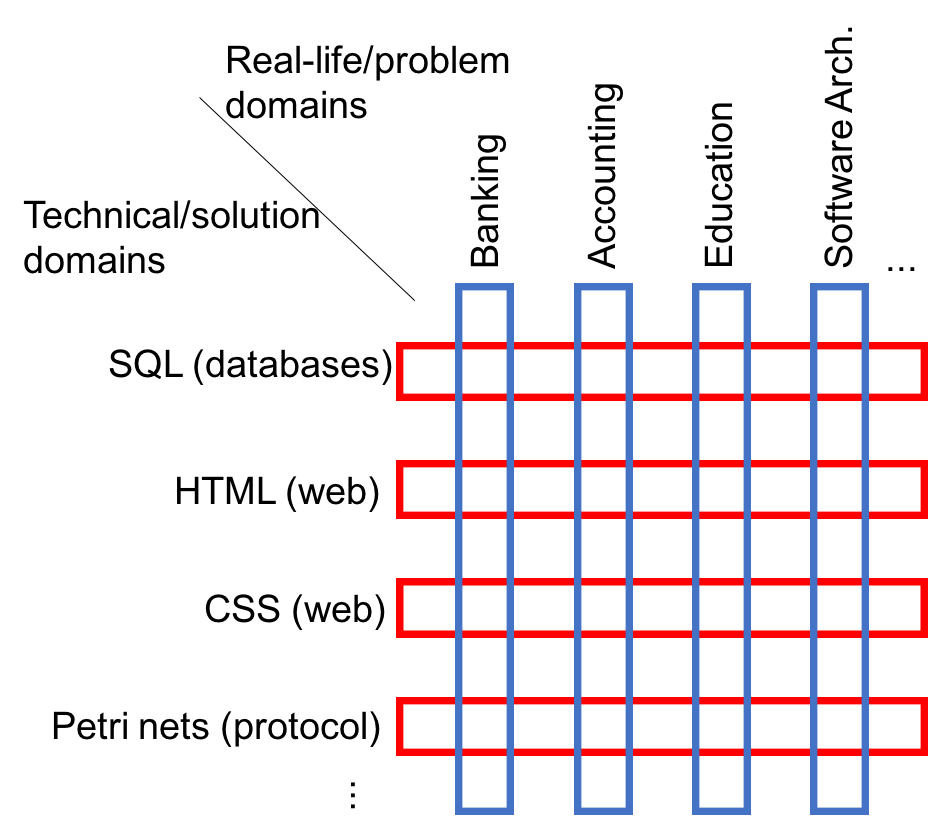}
	\caption{Solution and problem domains}
	\label{fig:solution_problem_domains}
\end{wrapfigure}

Domain-specific metamodelling is the art of creating and using languages which are specifically tailored for a particular domain~\cite{deLara2015,fowler2011dsl}.
In this case, the concept of ``domain'' corresponds to both real-life (problem) domains --- such as banking, accounting, robotics, home automation, cyber-physical systems, healthcare, product line systems, education, etc. --- and technical (solution) domains --- such as SQL, HTML, SysML, Petri Nets, Modelica, etc. --- (see Figure~\ref{fig:solution_problem_domains}).
The idea of DSMM is already practised in industry and academia and its positive results are documented in the literature~\cite{WhittleHR14,BurdenHW14,MohagheghiGSFNF13}.
However, mainstream approaches to DSMM and design and implementation of domain-specific languages (DSL), such as the Eclipse Modelling Framework~\cite{eclipse_modeling_framework_web}, the Unified Modelling Language, MetaCase~\cite{metacase}, etc., are based on two-level (meta)modelling approaches.
That is, domain concepts are defined in metamodels, and they are instantiated in models which conform to these metamodels.
This also includes the usage of ``profiles'' or ``stereotypes''.
Provided that in the real-life domains the way of thinking is not limited to a certain number of abstraction levels, approaches which force designers to adapt their way of thinking are deemed to fail.
Furthermore, describing software domains only in terms of models and their instances usually leads to unnecessary complexity and synthetic type-instance relations~\cite{atkinson2008reducing}.
In this work, we propose an alternative approach based on multilevel modelling (MLM) for DSMM, so that the number of abstraction levels are not limited.

MLM is already a recognised research area with clear advantages in several scenarios~\cite{delara2014whenandhow,atkinson2016melanee}.
However, it has several challenges that hamper its wide-range adoption, such as lack of a clear consensus on fundamental concepts of the paradigm, which has in turn led to lack of common focus in current multilevel tools~\cite{2015multi,atkinson2016multi}.
We tackle these challenges by combining the best of the worlds of two-level and multilevel (meta)modelling approaches~\cite{macias2016multecore}.
In our approach, each metamodel is a graph, and they are organised in a hierarchy representing natural abstraction levels.
Then, we define the advanced yet flexible relations between these graphs so that a (meta)model can be used to define concepts which are instantiated (and hence reused) on any lower levels in the hierarchy.
This is already a well-known technique in MLM, called \emph{deep characterisation}, and is usually achieved by means of the so-called \emph{potency}.
In this paper, we also revise the syntax and semantics of potency and use it in an advanced yet flexible manner.
We also provide a sound formal foundation for the approach based on graph transformations and category theory.

DSLs have already been used for the definition of both system's structure and behaviour~\cite{brambilla2012mdse,2015exe}.
Modelling structure has advanced due to mature tools and frameworks, and it is normally defined by a metamodel that determines the language concepts, the relationships between them, and the appropriate well-formedness constraints.
However, although understanding the behaviour of these models is required to understand the behaviour of the derived software systems, there is no clear consensus in the MDSE community when it comes to the specification of DSL behaviour.
Some approaches propose the use of UML behavioural models to represent system dynamics, such as in~\cite{EngelsUMLSemantics,StoryDiagrams,MayerhoferLWK13};
others make use of Abstract State Machines for the same purpose, like in~\cite{KM3formalization,Chen05};
and several authors define their own language to represent the behaviour of metamodels, such as the Kermeta language~\cite{KermetaWeaving}.
Nevertheless, we argue that in MDSE, where DSLs and model transformations are the key artifacts, model transformations are the natural way to specify behaviour. In particular, model transformation languages that support in-place update are very good candidates for the job~\cite{Rozenberg:1997,ehrig2006fundamentals}.

Understanding the behaviour of these models is required to understand the behaviour of the derived software systems.
However, there are several challenges regarding DSLs' behaviour, most of them related to the definition of their semantics.
Several approaches have been proposed for the definition and simulation of behavioural models based on model transformations (see, e.g.,~\cite{CsertanHMPPV02,LaraV02,taentzer2003agg,Rensink03,rivera2009graphical,Schurr2014}).
Although these approaches provide a very intuitive way to specify behavioural semantics, close to the language of the domain expert~\cite{JuanDeLara2006}, all of them rely on two-level modelling, which does not allow associating such behaviour to the right level of abstraction.

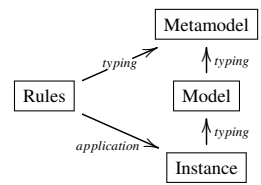
\begin{wrapfigure}[10]{r}[0mm]{.45\columnwidth}
	\scriptsize
	\centering
	$\xymatrix@R=3.3mm{
		& \framebox{Metamodel} \\
		\framebox{Rules} \ar[ur]|-{typing} \ar[dr]_{application} &
		\framebox{Model} \ar[u]_{typing} \\
		& \framebox{Instance} \ar[u]_{typing}
	}$
	\caption{Transformation rules describing semantics}
	\label{fig:transformation_two_levels_below}
\end{wrapfigure}

Many DSLs' metamodels and behaviours share significant commonalities, and hence being able to reuse model transformations across these DSLs would mean a significant gain.
Nevertheless, existing reusable model transformation approaches focus on traditional two-level modelling hierarchies (see, e.g.,~\cite{sanchez2008approaches,sanchez2011generic,delara2014flexiblereuse,duran2015amalgamation,chechik2016perspectives}, or~\cite{kusel2015reuse,kusel2013reality} for a survey).
This happens despite the fact that modelling behaviour is inherently multilevel since the behavioural modelling language is defined by metamodel while the semantics is described two levels below the metamodel~\cite{deLara2015,delara2010mixin,Rutle:2012:MAB}.
The cause is that behaviour is reflected in the running instances of the models which in turn conform to their metamodel (see Figure~\ref{fig:transformation_two_levels_below}).
Hence, MLM hierarchies could be used for DSMM, i.e., for the flexible organization of metamodelling languages, models and their running instances~\cite{deLara2015}.
Unfortunately, multilevel model transformations~\cite{atkinson2015enhancing,atkinson2012towards,atkinson2015execution} are relatively new and have not yet been proven suitable for both reuse and definition of behavioural models.
Writing a multilevel rule which refers to types on a higher level of abstraction will provide for higher degree of reusability and genericness, but the ``jump'' over the intermediate levels would lead to insufficient precision in the rules, which in turn might lead to too generic rules which would apply in undesired situations.

To achieve reusable multilevel model transformations, we propose the use of Multilevel Coupled Model Transformations (MCMTs) for the definition of behaviour: \textit{multilevel} to support the inherent multilevelness of domains and achieve reusability by genericness, and \textit{coupled} to support precision in rule definition and avoid repetition of very similar rules.
Hence, by utilising MLM for DSMM we could exploit commonalities among DSLs through abstraction, genericness and definition of behaviour by reusable model transformations.

We will use well-known terminology and constructions from category theory to formalise these multilevel hierarchies and their corresponding MCMTs.
By constructing a category for multilevel hierarchies and MCMTs, we are able to build upon the already existing co-span approach from the field of graph transformations~\cite{ehrig2009alternative}.
Moreover, it provides a precise semantics and the right intuition on the DSL definitions.
This paper's contributions can be beneficiary for two groups of readers: 
it could be used by category theory experts as a case-study on applying categorical constructions, in addition, 
it could be used by model transformation and DSL experts as a tool for the definition of reusable behaviour.
The former group of readers may skip Section~\ref{sec:tooling} while the latter group may skip Section~\ref{sec:chain-category}.

This paper is organised as follows.
In Section~\ref{sec:hierarchies}, we present the formal foundations of our modelling framework for building multilevel hierarchies of models, with a running example from the domain of Production Line Systems.
In Section~\ref{sec:mcmt} the proposed approach of multilevel, coupled model transformations is motivated and presented with the same running example.
The formal constructions behind this approach are outlined in Section~\ref{sec:chain-category}.
The tools and algorithms supporting the approach are presented in Section~\ref{sec:tooling}.
Finally, in Sections~\ref{sec:related-work} and~\ref{sec:conclusions-and-future-work} we present some related work, make some concluding remarks, and draw some directions for future work.

\section{Multilevel Metamodelling Hierarchies}
\label{sec:hierarchies}

The building blocks of a modelling framework are models.
Our models are represented by means of graphs, organised in a hierarchical manner.
In this section, we introduce the kind of graphs used to represent our models, as well as the way to organize them in hierarchies, by means of relations among graphs and the elements --- nodes and arrows --- that they contain.
We use as running example one from the domain of Product Line Systems (PLS), inspired by a common example used in several works for describing DSLs (see, e.g.,~\cite{CabotCGL08,rivera2009graphical}).
Our example hierarchy will progressively be completed in successive diagrams as we introduce the different elements of our proposal.
The diagram with the full hierarchy is displayed in Figure~\ref{fig:pls-multilevel-hierarchy} (Section~\ref{sec:pls-multilevel-hierarchy}), where we also explain in more detail the concepts defined in the graphs.

\subsection{Directed multigraphs}
\label{subsec:directed-multigraphs}

\begin{wrapfigure}[6]{r}[0mm]{.35\textwidth}
	\centering
	\vspace{-5mm}
	\includegraphics[page=1,width=\linewidth]{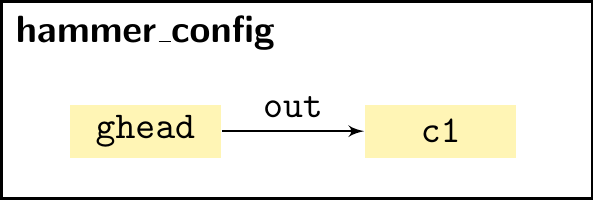}
	\caption{Directed multigraph with named elements}
	\label{fig:hierarchy-introductory-example-1}
\end{wrapfigure}

Our multilevel metamodelling approach is based on a flexible typing mechanism.
We will consider models abstractly as \emph{graphs} --- specifically, we work with directed multigraphs --- represented with a name, usually \graphname{G}{}.
These graphs consist of nodes and arrows.
A node, in the object-oriented modelling world, represents a class, and an arrow represents a relation between two classes.
Hence, an arrow always connects two nodes, and any two nodes can be connected by an arbitrary number of arrows.
Arrows with source and target in the same node (loops) are also allowed, and a node can likewise have any number of loops.
We will use the word \emph{element} to refer to both nodes and arrows, and assume that all elements are named and identified by such names.
For this reason, the names of any two nodes in the same graph must be different.
For the arrows, we allow for equal names as long as the source and target nodes do not match also, in order to be able to differentiate them.
Hence, arrows are triples \((s,a,t)\) with \(a\) an arrow name and $s$ and $t$ the names of its source and target nodes, respectively.
For a given graph \graphname{G}{}, we denote by \graphname[N]{G}{} its set of nodes, and by \graphname[A]{G}{} its set of arrows.
Also, a graph requires two mappings
\({\source[G]:\graphname[A]{G}{}\to\graphname[N]{G}{}}\)
and
\({\target[G]:\graphname[A]{G}{}\to\graphname[N]{G}{}}\)
that assign to each arrow its source and target nodes, respectively.
These two morphisms must be total to consider \graphname{G}{} a valid graph.
That is, we can define a graph as a quadruple
\(\graphname{G}{}{}={(\graphname[N]{G}{},\graphname[A]{G}{},\source[G],\target[G])}\).
Figure~\ref{fig:hierarchy-introductory-example-1} shows a graph named \elementname{hammer\_config}, which contains two nodes, \elementname{ghead} and \elementname{c1}, represented as yellow squares.
The arrow connecting these nodes is labelled with its name, \elementname{out}.

The relations between graphs, like typing and matching, are defined by \emph{graph homomorphisms}.
A homomorphism \(\phi\) from graph \graphname{G}{} to graph \graphname{H}{} is given by two maps
\({\phi^N:\graphname[N]{G}{}\to\graphname[N]{H}{}}\)
and
\({\phi^A:\graphname[A]{G}{}\to\graphname[A]{H}{}}\)
, compatible with the involved source and target maps, respectively.

Since graphs are used as the underlying representation of models, terms \textit{graph} and \textit{model} could be used interchangeably in this paper.
We will however differentiate by using the former for the definitions in this section, and the latter when discussing examples in the next ones.

\subsection{Tree shaped hierarchies, abstraction levels and typing chains}
\label{subsec:tree-shape-levels-and-tc}

We assume that a graph structure (model) is organized in a tree-shaped hierarchy with a single root graph.
Implicit in that assumption is the fact that each graph, except the one at the root, has exactly one parent graph in the hierarchy.
Also, we allow for arbitrary finite branching in the tree, so that each graph can have none or arbitrary finite many sibling graphs.

The hierarchy has \(l+1\) \emph{abstraction levels}, where \(\level{l}\) is the maximal length of paths in the hierarchy tree.
Each level in the hierarchy represents a different degree of abstraction.
Levels are indexed with increasing integers starting from the uppermost one, with index \(\level{0}\).
Each graph in the hierarchy is placed at some level \(\level{i}\), where \(\level{i}\) is the length of the path from the root to the graph.
We will use the notation \graphname{G}{i} to indicate that a graph is placed at level \(\level{i}\).
Level~\(\level{0}\) contains, in such a way, just the root \graphname{G}{0} of the tree.%
\footnote{For implementation reasons, we always use Ecore~\cite{steinberg2008emf} as root graph at level \(\level{0}\), although we will not fully display it in this section, and will omit it completely in the following ones.}
For any graph \graphname{G}{i} we call the unique path \(\tc{\graphname{G}{i}} = [\graphname{G}{i}, \graphname{G}{i-1}, \dots, \graphname{G}{1}, \graphname{G}{0}]\) from this graph to the root graph of the hierarchy the \emph{typing chain} of \graphname{G}{i}.

\begin{figure}
	\centering
	\includegraphics[page=2,width=.6\linewidth]{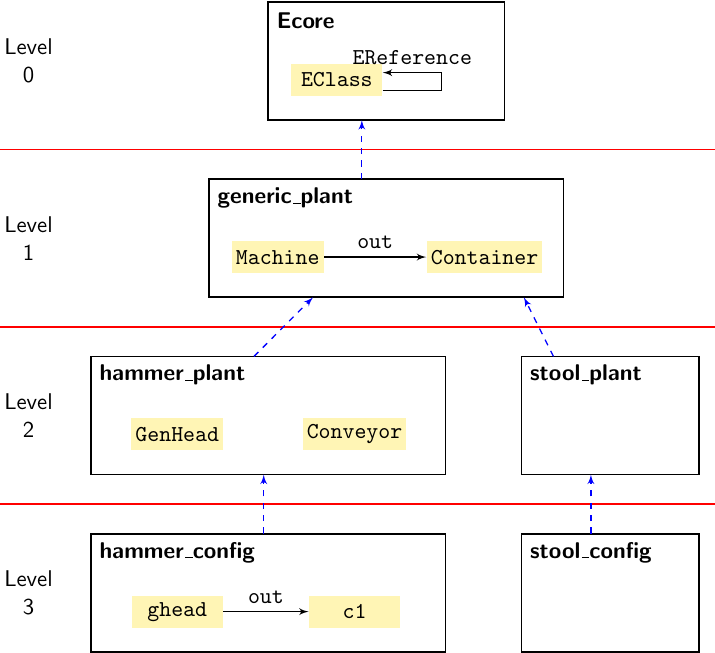}
	\caption{Tree-shaped hierarchy of graphs with levels and typing chains}
	\label{fig:hierarchy-introductory-example-2}
\end{figure}

In Figure~\ref{fig:hierarchy-introductory-example-2} we show several graphs that constitute a tree-shaped hierarchy, included the one in Figure~\ref{fig:hierarchy-introductory-example-1}.
We can see the graph \elementname{hammer\_config}, representing a particular configuration of the elements of some hammer factory, in level \(\level{3}\), together with another graph \elementname{stool\_config}, defining a different configuration for a specific stool-producing plant.
These two graphs, although conceptually similar, are independent from each other in the sense that they neither belong to the same branch of the tree nor share their parent graphs --- that is, their metamodels.
Since we focus in this paper in behavioural modelling languages, the models we define evolve through time, representing the execution of the modelled system.
Thus, the purpose of these two models is not just representing a specific plant for producing hammers or stools.
They also store, at any particular point in time, the state of the execution of the system.
That is, the way machines produce and modify parts in order to manufacture the finished hammers or stools.

\subsection{Individual typing}
\label{subsec:individual-typing}

In the hierarchy in Figure~\ref{fig:hierarchy-introductory-example-2}, the parent graphs of \elementname{hammer\_config} and \elementname{stool\_config} are located at level \(\level{2}\), where most of the types for the elements of the previous two graphs are defined.
These two parent graphs, \elementname{hammer\_plant} and \elementname{stool\_plant}, define the types of elements that we can use to create a specific setting of a plant for producing each type of object.
Both models share the common parent graph \elementname{generic\_plant}, located at level~\(\level{1}\).
This language contains basic, abstract concepts for plant configurations, like \elementname{Machine} and \elementname{Container}.
Finally, as we previously mentioned, we can see \elementname{Ecore}, the topmost graph and root of the tree, at level \(\level{0}\).
This graph is the parent graph of \elementname{generic\_plant}.
Note that we use red horizontal lines to indicate the separation between two levels, and blue dashed arrows indicating the sequences of graphs that constitute typing chains and provide the required tree shape.
Note also that, since the purpose of this figure is to illustrate the tree shape of the metalevel hierarchies, the contents of both graphs \elementname{stool\_plant} and \elementname{stool\_config} are not displayed.
The diagram with the full hierarchy is displayed in Figure~\ref{fig:pls-multilevel-hierarchy}.

Any element \elementname{e} in any graph \graphname{G}{i} has a unique type denoted by \(\typeof{\elementname{e}}\).
In that case, we can say that \elementname{e} is \emph{typed by} \(\typeof{\elementname{e}}\) or, equivalently, that \elementname{e} is an \emph{instance} of \(\typeof{\elementname{e}}\).
Inheritance relations among nodes in the same level are allowed, although they are not used in the examples in this paper.
The semantics of this relation are that the information of the parent node (potency and typing) and all its attributes and relations --- both incoming and outgoing --- are replicated in the child node.
That is, if a node \elementname{x} has an inheritance relation to a node \elementname{y} of type \elementname{Y} (i.e. \elementname{y} is a parent node of \elementname{x}), then \elementname{x} must necessarily be typed by \elementname{Y}, and have the same potency, attributes and relations as \elementname{y}.
Inheritance relations cannot be established between nodes in different models or be cyclic, neither directly (self-inheritance) or transitively.

To achieve the necessary flexibility, we allow typing to jump over levels.
That is, for any \elementname{e} in a graph \graphname{G}{i}, with \(i\geq 1\), its \emph{individual (or direct) type} \(\typeof{\elementname{e}}\) is found in a graph \(\typegraphof{\elementname{e}}\), which is one of the graphs \(\graphname{G}{i-1}, \dots, \graphname{G}{1}, \graphname{G}{0}\) in its typing chain.
Note that the graphs in which we locate the types of different elements in \graphname{G}{i} may also be different.
By \(\difference{\elementname{e}}\) we denote the difference between \(\level{i}\) and the level where \(\typegraphof{\elementname{e}}\) is located.
In most cases, this difference is \(1\), meaning that the type of \elementname{e} is located at the level directly above it.
In short, we can say that, for any element \elementname{e} in a given graph \graphname{G}{i}, with \(i\geq 1\), its type \(\typeof{\elementname{e}}\) is an element in the graph \(\graphname{G}{i-\difference{\elementname{e}}}\), where \(1\leq \difference{\elementname{e}}\leq i\).

\begin{figure}
	\centering
	\includegraphics[page=3,width=.6\linewidth]{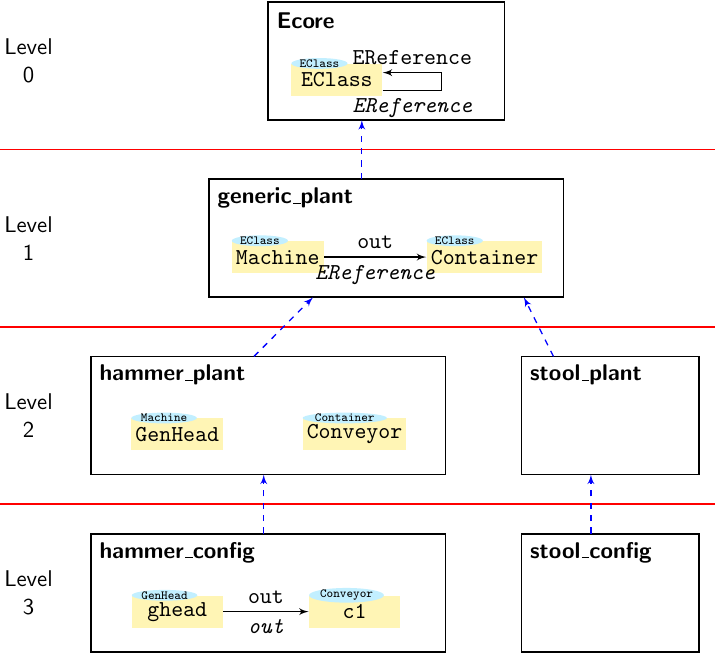}
	\caption{Graph hierarchy with typed elements}
	\label{fig:hierarchy-introductory-example-3}
\end{figure}

Figure~\ref{fig:hierarchy-introductory-example-3} displays our example hierarchy again, including in this case the type of each element.
To avoid polluting the diagram with too many arrows, we use alternative representations for them.
For every node, its type is identified by name and depicted in a blue ellipse attached to the node.
For example, the type of \elementname{GenHead} in \elementname{hammer\_plant} is \elementname{Machine}, a relation that could also be represented as an arrow between these two nodes.
For the actual arrows that represent relations among nodes in the same level, the type is represented as a second label with the name of the type in italics font.
For example, the type of the \elementname{out} arrow in \elementname{generic\_plant} is \elementname{EReference}, located in the \elementname{Ecore} graph.

To ensure that every element has a type, we assume that the root graph \graphname{G}{0} has a \emph{self-defining} collection of individual typing assignments.%
\footnote{
We use the more relaxed term \emph{self-defining} instead of \emph{reflexive} since the former requires all the elements in the graph to be typed by elements in the same graph, whereas the latter requires every element to be strictly typed by itself.
That is, instead of requiring that \(\typeof{\elementname{e}}=\elementname{e}\) for all \(\elementname{e}\in\graphname{G}{0}\), we just need to ensure that if \(\elementname{e}\in\graphname{G}{0}\) then \(\typeof{\elementname{e}}\in\graphname{G}{0}\).
}
Possible candidates for the root graph \graphname{G}{0} are, for instance, Ecore --- the one we chose --- and the node-and-arrow graph described in~\cite{rutle2012formal}.

From a more general point of view, we obtain for any \elementname{e} in \graphname{G}{i} a sequence
$$
\xymatrix@R=7mm{
\elementname{e}
  \ar@{|->}[r]
  \ar@{}[d]|{\rotatebox[origin=c]{-90}{\(\in\)}}
& \typeof{\elementname{e}}
  \ar@{|->}[r]
  \ar@{}[d]|{\rotatebox[origin=c]{-90}{\(\in\)}}
& \typeof[2]{\elementname{e}}
  \ar@{}[d]|{\rotatebox[origin=c]{-90}{\(\in\)}}
  \ar@{|->}[r]
& \dots
  \ar@{|->}[r]
& \typeof[\step{\elementname{e}}]{\elementname{e}}
  \ar@{}[d]|{\rotatebox[origin=c]{-90}{\(\in\)}} \\
\graphname{G}{i}
& \graphname{G}{i-\difference{\elementname{e}}}
& \graphname{G}{i-{\difference[2]{\elementname{e}}}}
&& \graphname{G}{i-{\difference[\step{\elementname{e}}]{\elementname{e}}}}
  \ar@{}[r]|-{=}
& \graphname{G}{0}
}
$$
of typing assignments of length \(1\le\step{\elementname{e}}\le{i}\) with \((i-{\difference[\step{\elementname{e}}]{\elementname{e}}})=0\).
The number \(\step{\elementname{e}}\) of steps depends individually on the item \elementname{e}.
For convenience, we use the following abbreviations:

\vspace{-3mm}
\begin{equation*}
\begin{aligned}[t]
\typeof[2]{\elementname{e}}&=\typeof{\typeof{\elementname{e}}}\\
\difference[2]{\elementname{e}}&=\difference{\elementname{e}}+\difference{\typeof{\elementname{e}}}\\
\end{aligned}
\hspace{1cm}
\begin{aligned}[t]
\typeof[3]{\elementname{e}}&=\typeof{\typeof{\typeof{\elementname{e}}}}\\
\difference[3]{\elementname{e}}&=\difference[2]{\elementname{e}}+\difference{\typeof[2]{\elementname{e}}}\\
\end{aligned}
\begin{aligned}[t]
\dots\\
\dots\\
\end{aligned}
\end{equation*}
We call any of the elements $\typeof{\elementname{e}}$, $\typeof[2]{\elementname{e}}$, $\typeof[3]{\elementname{e}}$, \ldots a \emph{transitive type} of $\elementname{e}$.

Let us consider an arbitrary arrow \(\elementname{x}\xrightarrow{\elementname{f}}\elementname{y}\), together with its source and target nodes in a graph \graphname{G}{i}.
The types of the three elements may be located in three different graphs.
The typing of arrows should, however, be compatible with the typing of sources and targets, that is, the source and the target of an arrow \(\typeof{\elementname{f}}\in\graphname{G}{i-\difference{\elementname{f}}}\) must be provided by the types of \(\elementname{x}\) and \(\elementname{y}\), respectively:

$$
\xymatrix@R=7mm{
\graphname{G}{i}
  \ar@{}[d]|{\rotatebox[origin=c]{90}{\(\in\)}}
& \graphname{G}{i-\difference{\elementname{x}}}
  \ar@{}[d]|{\rotatebox[origin=c]{90}{\(\in\)}}
& \graphname{G}{i-\difference{\elementname{y}}}
  \ar@{}[d]|{\rotatebox[origin=c]{90}{\(\in\)}}
& \dots
& \graphname{G}{i-\difference{\elementname{f}}}
  \ar@{}[d]|{\rotatebox[origin=c]{90}{\(\in\)}} \\
\elementname{x}
  \ar@{|->}[r]
  \ar[d]^{\elementname{f}}
& \typeof{\elementname{x}}
  \ar@{|->}[rr]
&& \dots
  \ar@{|->}[r]
& \typeof[m_\elementname{x}]{\elementname{x}}
  \ar[d]^{\typeof{\elementname{f}}}             \\
\elementname{y}
  \ar@{|->}[rr]
&& \typeof{\elementname{y}}
  \ar@{|->}[r]
& \dots
  \ar@{|->}[r]
& \typeof[m_\elementname{y}]{\elementname{y}}
}
$$

\noindent More precisely, we require that this \emph{non-dangling typing} condition is satisfied:
There exist \(1\le m_\elementname{x}\le\step{\elementname{x}}\) and \(1\le m_\elementname{y}\le\step{\elementname{y}}\) such that \(\difference[m_\elementname{x}]{\elementname{x}}= \difference[m_\elementname{y}]{\elementname{y}}=\difference[]{\elementname{f}}\), \(\typeof[m_\elementname{x}]{\elementname{x}}\) is the source of \(\typeof{\elementname{f}}\), and \(\typeof[m_\elementname{y}]{\elementname{y}}\) is the target of \(\typeof{\elementname{f}}\).

\subsection{Typing morphisms and domains of definition}
\label{subsec:typing-morphisms-and-domain-definition}

Our notions of flexible multilevel typing can be described for graphs, in a more abstract and compact way, by means of a family of typing morphisms.
That is, we can define typing relations between any two graphs in a typing chain as an abstraction of the individual typing that we defined for their elements in Section~\ref{subsec:individual-typing}.
The vocabulary defined for individual typing can be reused here, so that a graph can be an \emph{instance} of another graph, or be \emph{typed by} it.
These relations among graphs are defined by means of graph homomorphisms.
Since the individual typing maps are allowed to jump over levels, two different elements in the same graph may have their types located in different graphs along the typing chain.
Hence, the typing morphisms established between graphs become partial graph homomorphisms~\cite{rossini2014formalisation}.

\begin{wrapfigure}[18]{r}[0mm]{.4\textwidth}
	\centering
		\includegraphics{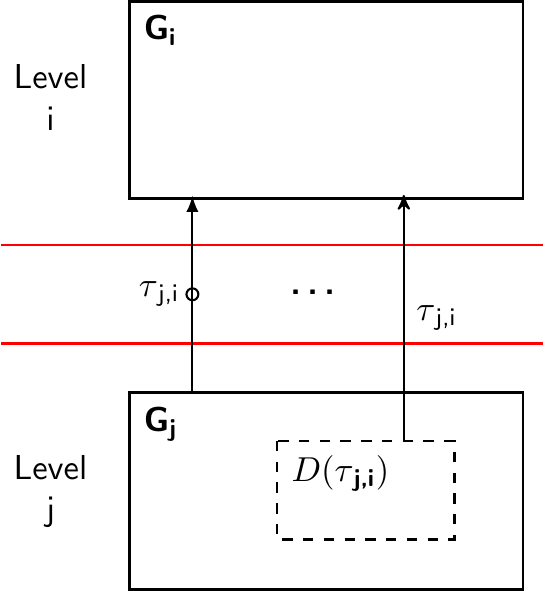}
		\caption{Typing morphisms and domain of definition}
		\label{fig:tau-and-domain-of-definition}
\end{wrapfigure}

Our characterization of individual typing ensures that, for any levels \(\level{i}\), \(\level{j}\) such that \(0\le\level{i}<\level{j}\le\level{l}\), there is a \emph{partial typing morphism} \(\typemorph{j}{i}:\graphname{G}{j}\partialmap\graphname{G}{i}\) given by a subgraph \(\domain{\typemorph{j}{i}}\sqsubseteq\graphname{G}{j}\), called the \emph{domain of definition} of \(\typemorph{j}{i}\), and a \emph{total typing homomorphism} \(\typemorph{j}{i}:\domain{\typemorph{j}{i}}\to\graphname{G}{i}\).
The domain of definition may be empty --- and, consequently, \(\typemorph{j}{i}\) is just the inclusion of the empty graph in \(\graphname{G}{i}\) --- in case no \(\elementname{e}\in\graphname{G}{j}\) has a transitive type \(\typeof[m]{\elementname{e}}\in\graphname{G}{i}\) for some $m\geq 1$.
Also, in abuse of notation, we use the same name for both morphisms, since they represent the same typing information.
Using the same syntax as the examples, Figure~\ref{fig:tau-and-domain-of-definition} depicts these concepts in a generic manner.

Given that a graph is given by its set of nodes and its set of arrows, we can define the domain of definition component-wise.
For any \(0\le\level{i}<\level{j}\le\level{l}\) we can define the set \(\domain{\typemorph[N]{j}{i}}\subseteq\graphname[N]{G}{j}\) of all nodes in \(\graphname{G}{j}\) that are transitively typed by nodes in \(\graphname{G}{i}\).
That is, \(\elementname{e}\in\domain{\typemorph[N]{j}{i}}\) iff there exists \(1\le m\le\step{\elementname{e}}\) such that \(\level{j}-\level{i}=\difference[m]{\elementname{e}}\), and thus \(\typeof[m]{\elementname{e}}\in\graphname[N]{G}{i}\).
We set \(\typemorph[N]{j}{i}(\elementname{e}):=\typeof[m]{\elementname{e}}\) for all nodes in \(\domain{\typemorph[N]{j}{i}}\) and obtain, in such a way, a map \(\typemorph[N]{j}{i}:\domain{\typemorph[N]{j}{i}}\to\graphname[N]{G}{i}\).
This total map defines a partial map \(\typemorph[N]{j}{i}:\graphname[N]{G}{j}\partialmap\graphname[N]{G}{i}\) with the domain of definition \(\domain{\typemorph[N]{j}{i}}\).
This component-wise construction is required to ensure the compatibility between the way that domains of definition are calculated and the individual typing.

We can follow an analogous procedure for the set \(\graphname[A]{G}{}\) to obtain a partial map \(\typemorph[A]{j}{i}:\graphname[A]{G}{j}\partialmap\graphname[A]{G}{i}\) with domain of definition \(\domain{\typemorph[A]{j}{i}}\).

The non-dangling typing condition is now equivalent to the requirement that the pair \((\domain{\typemorph[N]{j}{i}},\domain{\typemorph[A]{j}{i}})\), constitutes a subgraph \(\domain{\typemorph{j}{i}}\) of \(\graphname{G}{j}\).
That is, for any arrow \(\elementname{e}\in\domain{\typemorph[A]{j}{i}}\) we have \(\source[\graphname{G}{j}](\elementname{e}),\target[\graphname{G}{j}](\elementname{e})\in\domain{\typemorph[N]{j}{i}}\).
Consequently, the pair of morphisms \((\typemorph[N]{j}{i},\typemorph[A]{j}{i})\) provides a total graph homomorphism \({\typemorph{j}{i}:\domain{\typemorph{j}{i}}\to\graphname{G}{i}}\) and thus a partial typing morphism \(\typemorph{j}{i}:\graphname{G}{j}\partialmap\graphname{G}{i}\).

All the typing morphisms \(\typemorph{j}{0}:\graphname{G}{j}\partialmap\graphname{G}{0}\) are total, since every element has a type.
The uniqueness of typing is reflected on the abstraction level of type morphisms by the \emph{uniqueness condition}: For all \(0\le\level{i}<\level{j}<\level{k}\leq\level{l}\), we have that \(\typemorph{k}{j};\typemorph{j}{i}\preceq\typemorph{k}{i}\), i.e., \(D(\tau_{k,j};\tau_{j,i})\sqsubseteq D(\tau_{k,i})\) and, moreover, \(\tau_{k,j};\tau_{j,i}\) and \(\tau_{k,i}\) coincide on \(D(\tau_{k,j};\tau_{j,i})\). Note, that the domain of definition of the composition \(D(\tau_{k,j};\tau_{j,i})\) is obtained by a pullback (inverse image) (see Definition~\ref{def:graph-chain} in Section~\ref{subsec:multilevel-typing}). Informally, the uniqueness condition means whenever an element in \graphname{G}{k} is transitively typed by an element in \graphname{G}{j} such that this element in \graphname{G}{j} is, in turn, transitively typed by an element in \graphname{G}{i} then the element in \graphname{G}{k} is also transitively typed by an element in \graphname{G}{i} and both ways provide the same type in \graphname{G}{i}.

The other way around, we can reconstruct individual typings from a family of partial typing morphisms between graphs that satisfy the totality and uniqueness conditions.
For any item \elementname{e} in a graph \graphname{G}{k} there exists a maximal (least abstract) level \(0\le\level{i_\elementname{e}}<\level{k}\) such that \elementname{e} is in \(\domain{\typemorph{k}{i_\elementname{e}}}\) but not in \(\domain{\typemorph{k}{j}}\) for all \({i_\elementname{e}}<\level{j}<\level{k}\), since \(\typemorph{j}{0}\) is total and \(\level{k}\) a finite number.
Hence, the individual type of \elementname{e} is given by \(\typeof{\elementname{e}}:=\typemorph{k}{i_\elementname{e}}(\elementname{e})\) and \(\difference{\elementname{e}}:=\level{k}-\level{i_\elementname{e}}\).

\begin{figure}
	\centering
	\includegraphics[page=4,width=.6\linewidth]{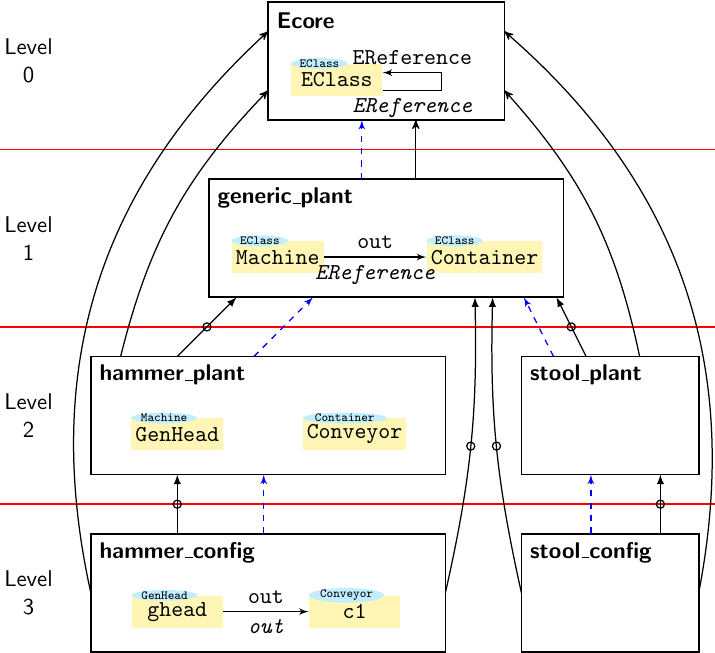}
	\caption{Graph hierarchy with typing morphisms}
	\label{fig:hierarchy-introductory-example-4}
\end{figure}

Figure~\ref{fig:hierarchy-introductory-example-4} depicts the typing morphisms between all the graphs in the hierarchy.
Note that the morphism from \elementname{generic\_plant} to \elementname{Ecore} is total, since all the types used in the former can only be defined in the latter, and all elements must have a type.

\subsection{Potency}
\label{subsec:potency}

In this subsection, we introduce our modification and formalisation of the concept of \emph{potency of an element} originally defined in~\cite{atkinson2002rearchitecting}.
Potencies are used on elements as a means of restricting the length of the jumps of their typing morphisms across several levels.
The reason for this is that the formalisation presented so far is aimed at classifying graphs in tree-shaped hierarchies as a means to get a clear structure of the concepts defined, but the hierarchy is built based on the individual typing relations, which do not have any restrictions regarding levels.
Due to this fact, levels become less useful in practice if the individual typing relations are unbounded, hence the necessity of potencies to restrict the jumps of typing across levels.
This concept is therefore designed to allow designers to properly constrain how nodes and arrows are instantiated in a more detailed way, but it is not required to ensure the correct typing of nodes or the typing compatibility of arrows.

We represent the potency of an element with an interval, using the notation \elementname{min-max}.
These values may appear after the declaration of an element using \elementname{\potencyseparator} as a separator.
More precisely, for any element \elementname{x} in \graphname{G}{i} we require \(min\le\difference{\elementname{x}}\le max\) where \typename{\(\typeof{\elementname{x}}\)}{min-max} is the declared potency of the type \(\typeof{\elementname{x}}\) in \(\graphname{G}{i-\difference{\elementname{x}}}\).
This condition can be reformulated using partial typing morphisms.
For any element \elementname{y} in \graphname{G}{j} with a potency declaration \typename{y}{min-max} we require that \(\typemorph[-1]{j}{i}(\elementname{y})\) is empty for all \(\level{j}\) with \(\level{j-i}<\elementname{min}\) or \(\level{j-i}>\elementname{max}\).
In the cases where \(\elementname{min} = \elementname{max}\), the notation can be simplified to show just one number.
Such is the case with the default value of potency: \(\typename{}{1-1} \equiv \typename{}{1}\).

\begin{figure}
	\centering
	\includegraphics[page=5,width=.6\linewidth]{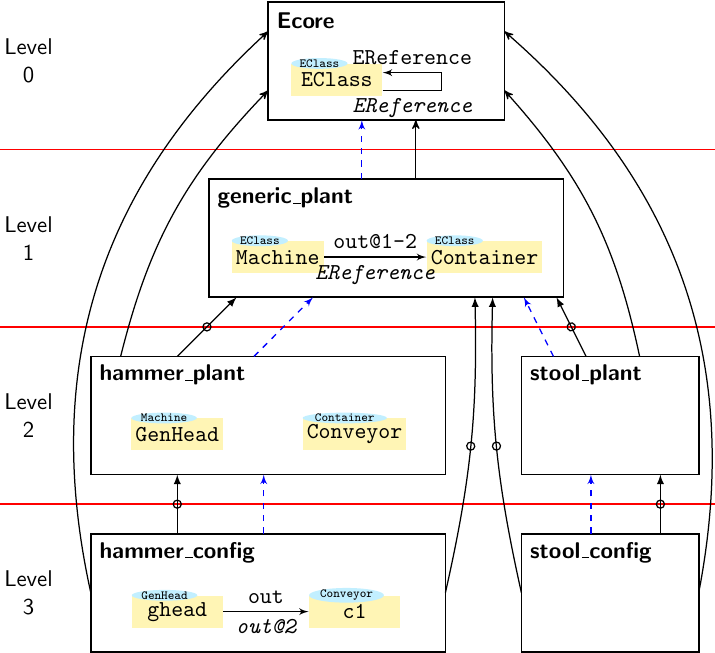}
	\caption{Full graph hierarchy with potency}
	\label{fig:hierarchy-introductory-example-5}
	\vspace{-4mm}
\end{figure}

For the typing relations used in our example to be correct, we require the arrow \elementname{out} in \elementname{generic\_plant} to have increased potency.
That way, it becomes possible to use it as type for the other \elementname{out} arrow, located \emph{two levels below}, in the \elementname{hammer\_config} graph.
Besides, and to ease the interpretation of a diagram, the type of an element which is specified in a level different from the one immediately above uses the \elementname{\potencyseparator} notation to indicate it.
For nodes, the annotation is located in the blue ellipse.
For arrows, the \typename{ty(e)}{\(\difference{\elementname{e}}\)} annotation is displayed in the declaration of the type (in italics).
Figure~\ref{fig:hierarchy-introductory-example-5}
shows a new version of the example hierarchy which includes potencies.

Unlike other approaches, our realization makes the declared potency of an element independent of the potency of its type.
For example, the default value of the potency of the arrow \elementname{out} in \elementname{hammer\_config} could be changed to some other one without being affected by the potency \typename{}{1-2} of its type arrow (\elementname{out} in \elementname{generic\_plant}).

\subsection{PLS multilevel hierarchy}
\label{sec:pls-multilevel-hierarchy}

\begin{figure}
	\centering
	\includegraphics[width=.8\textwidth]{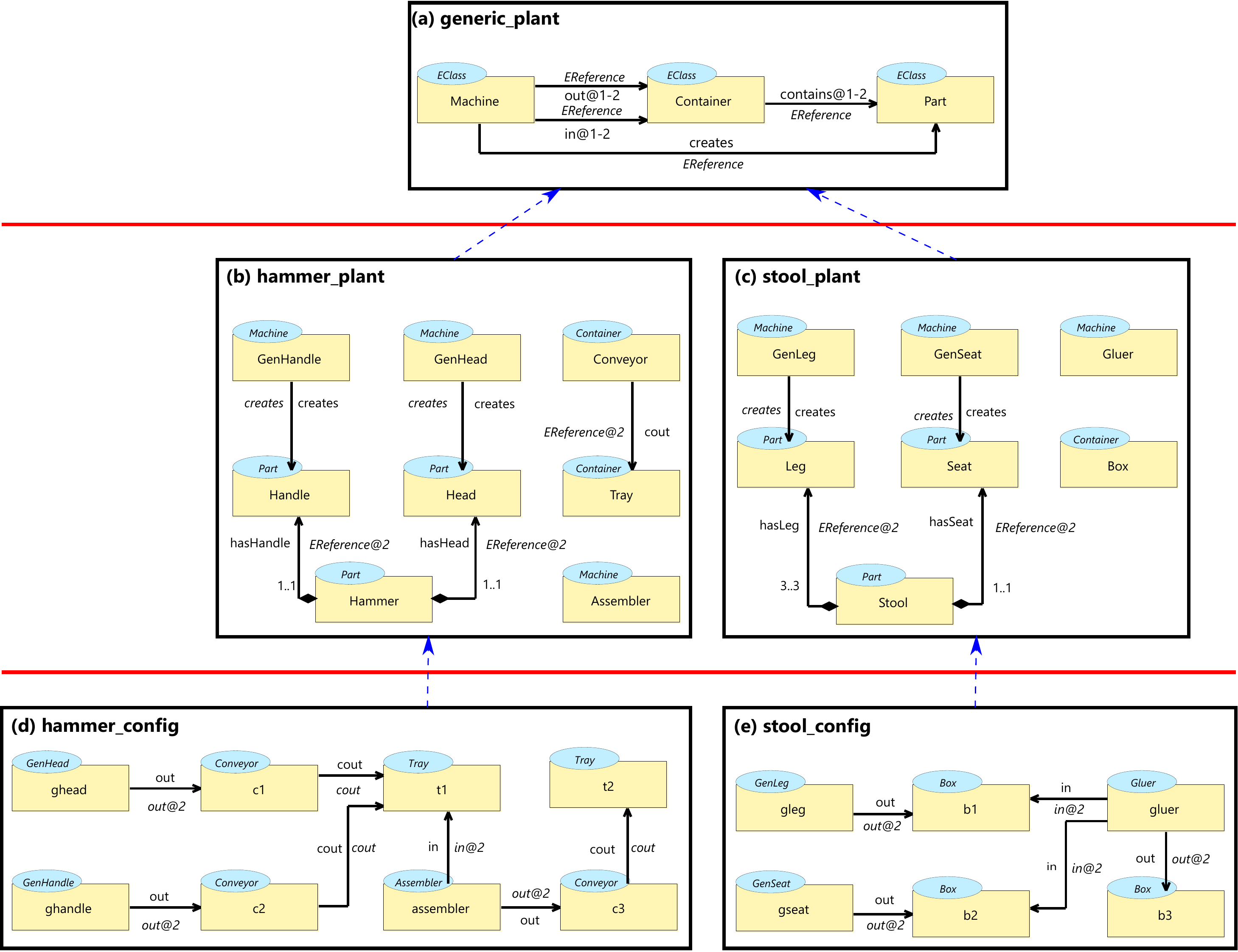}
	\caption{Full hierarchy for the PLS case study}
	\label{fig:pls-multilevel-hierarchy}
\end{figure}

The full version of the PLS hierarchy is shown in Figure~\ref{fig:pls-multilevel-hierarchy}.
This figure has been generated from the real implementation of the PLS multilevel hierarchy created with the tool MultEcore (see Section~\ref{sec:tooling}).
For this reason, there is an addition to the syntax used so far:
The multiplicity of arrows is now displayed in those cases where it is different from its default value \elementname{0..n}.

The hierarchy displays a top model, called \elementname{generic\_plant} (Figure~\ref{fig:pls-multilevel-hierarchy}a), where we define abstract concepts related to product lines that manufacture physical objects.
\elementname{Machine} defines any device that can create, modify or combine objects, which are represented by the concept \elementname{Part}.
In the first case, we indicate the relation from a generator-like machine to the part it generates with the \elementname{creates} relation.
In order for parts to be transported between machines or to be stored, they can be inside \elementname{Containers}, and this relation is expressed by the \elementname{contains} relation.
All machines may have containers to take parts from or to leave manufactured ones in.
These two relations are identified with the \elementname{in} and \elementname{out} relations, respectively.

The second level of the hierarchy contains two models, that define concepts related to specific types of plants.
In Figure~\ref{fig:pls-multilevel-hierarchy}b we can find \elementname{hammer\_plant}, where the final product \elementname{Hammer} is created by combining one \elementname{Handle} and one \elementname{Head}.
Both \elementname{hasHandle} and \elementname{hasHead} relations express this fact, and their \elementname{1..1} multiplicity states that a \elementname{Hammer} must be created out of exactly one \elementname{Head} and one \elementname{Handle}.
The type of these two relations is \elementname{EReference} (from \elementname{Ecore}), since there is no relation defined for parts in \elementname{generic\_plant}, because the concept of assembling parts is too specific to be located in \elementname{generic\_plant}.
Note that thanks to the use of potency, we can define \elementname{hasHandle} and \elementname{hasHead} without forcing \elementname{Part} to have a relation with itself in \elementname{generic\_plant} in order to use it as their type: It is not desirable that a specific detail in a lower level (\elementname{hammer\_plant}) affects the specification of an upper one (\elementname{generic\_plant}).
In the \elementname{hammer\_plant} model we also define three types of machines.
First, \elementname{GenHandle} and \elementname{GenHead}, that create the corresponding parts, indicated by the two \elementname{creates} arrows.
And secondly, \elementname{Assembler}, that generates hammers by assembling the corresponding parts.
Finally, the model contains two specific instances of \elementname{Container}, namely \elementname{Conveyor} and \elementname{Tray}.
The \elementname{cout} arrow between them indicates that a \elementname{Conveyor} must always be connected to a \elementname{Tray}.
As explained before, this relation can be defined without requiring a loop arrow in \elementname{Container}, thanks to the use of potency, satisfying the separation in levels of abstraction.

The other model in the second level, depicted in Figure~\ref{fig:pls-multilevel-hierarchy}c, contains another specification of a manufacturing plant, in this case for stools.
In this model, the relations between \elementname{Stool}, \elementname{Leg} and \elementname{Seat} resembles those of \elementname{Hammer}, \elementname{Handle} and \elementname{Head}.
The multiplicity of the arrow between the first two indicates that a \elementname{Stool} must have exactly three \elementname{Legs}.
Two different types of machine, \elementname{GenLeg} and \elementname{GenSeat}, manufacture \elementname{Legs} and \elementname{Seats}, respectively.
The remaining one, \elementname{Gluer}, takes three \elementname{Legs} and a \elementname{Seat} and creates a \elementname{Stool} out of them.
Finally, the only container defined for this kind of plant is \elementname{Box}.

The two models at the bottom of the hierarchy, in Figures~\ref{fig:pls-multilevel-hierarchy}d and~\ref{fig:pls-multilevel-hierarchy}e, represent specific configurations of a hammer PLS (\elementname{hammer\_config}) and a stool PLS (\elementname{stool\_config}).
They contain specific instances of the concepts defined in the level above, organized to specify correct product lines, in which parts get transferred from generator machines to machines that combine them to obtain the final manufactured products.
Note that, again, the use of potencies enables the creation of instances of the \elementname{out} relation defined two levels above between \elementname{Machine} and \elementname{Container}, without requiring a type defined in the intermediate level of \elementname{hammer\_plant} and \elementname{stool\_plant}.

\section{Multilevel Coupled Model Transformations for the Definition of Behaviour}
\label{sec:mcmt}

To enable the integration of MLM in MDSE projects, model-to-model transformations are key.
In the literature, we find notions such as multilevel to two-level model transformations~\cite{atkinson2012towards}, deep transformations --- defined at a specific level but with references to upper levels ---~\cite{deLara2015}, or coupled transformations --- that operate on a model and its metalevels simultaneously~\cite{HerrmannsdorferW14,deLara2015,Rutle:2012:MAB}.
In this section, we propose a generic way of specifying multilevel transformations, operating on models that belong to a multilevel hierarchy.
We focus on in-place transformations, and specifically use them to define the behaviour of DSLs.

Indeed, the formal, flexible MLM approach introduced so far also opens a door for the reusability of model's behaviour.
Since most behavioural models have some commonalities both in concepts and their semantics, reusing these model transformations across behavioural models would mean a significant gain.
By utilising MLM in a metamodelling process for the definition of modelling languages, we can exploit commonalities among these languages through abstraction, genericness and reusability of behaviour.
We will build on our running example from the PLS domain to explain our approach to reusable model transformations, namely, \emph{multilevel coupled model transformations} (MCMTs).
We will also compare MCMTs to other well-known approaches so that the advantages become clearer.

\subsection{Why using MCMTs?}
\label{subsec:why-using-mcmt}

Recall the PLS hierarchy above, and assume we want to define the behaviour of the machines that generate parts.
Instances of \elementname{Machine} are related by the \elementname{creates} relation with instances of \elementname{Part}.
Using model transformations, we have now three options to define the rule for the action of creating a part, namely, traditional two-level model transformation rules, multilevel rules, and our multilevel coupled model transformations.
For the display of the example rules in the following, we use a declarative way of specifying transformation rules, and we do this using the same graphical syntax for nodes, arrows, types and potencies that we used in Section~\ref{sec:hierarchies}.

\subsubsection{Traditional two-level model transformation rules}
\begin{figure}
    \begin{minipage}{0.45\textwidth}
		\centering
		\includegraphics[scale=.75]{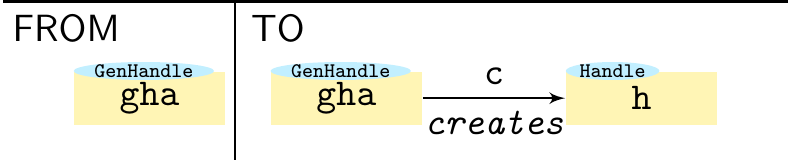}
		\caption{Behaviour for the \elementname{GenHandle} machine}
		\label{fig:pls-rule-create-part-2-level-1}
    \end{minipage}\hfill
    \begin{minipage}{0.45\textwidth}
		\centering
		\includegraphics[scale=.75]{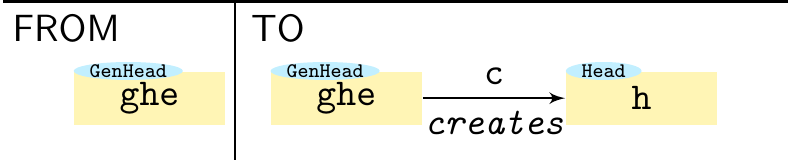}
		\caption{Behaviour for the \elementname{GenHead} machine}
		\label{fig:pls-rule-create-part-2-level-2}
	\end{minipage}
\end{figure}

Using two-level transformation rules for the creation of parts on instances of \elementname{hammer\_plant} (Figure~\ref{fig:pls-multilevel-hierarchy}b), we would need to specify two rules, one for each generator.
That is, one for \elementname{GenHandle} (Figure~\ref{fig:pls-rule-create-part-2-level-1}) and one for \elementname{GenHead} (Figure~\ref{fig:pls-rule-create-part-2-level-2}).
If we wanted to do the same for the stool plant (Figure~\ref{fig:pls-multilevel-hierarchy}c), we would need two more rules.

Following the declarative specification of transformation rules, the rules consist of FROM and TO blocks. 
The FROM block contains the pattern that the rule must match in order to be executed.
If a match is found, then the matched subgraph is modified in the way defined in the TO block, which may comprise the creation, deletion or modification of nodes and arrows.

While this might do the job for one language, problems arise regarding reusability.
The rules would be too specific and tied to the types \elementname{GenHandle}, \elementname{GenHead}, \elementname{Handle} and \elementname{Head}, and the creation of further parts in this or other languages would require additional specific rules.
As a consequence, a significant number of very similar rules are required, leading to the \emph{proliferation problem}: each machine would need a rule and each hierarchy would need its set of almost \textit{identical} rules.

\begin{wrapfigure}[12]{r}[0mm]{.5\textwidth}
	\centering
	\includegraphics{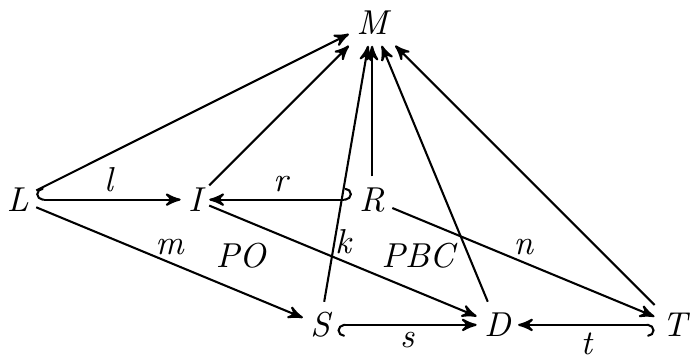}
	\caption{Two-level MT rule}
	\label{fig:two-level-rule-formal}
\end{wrapfigure}

The basic structure of these rules is outlined in Figure~\ref{fig:two-level-rule-formal}.
In its most general terms, a graph transformation rule is defined as a left \(L\) and a right \(R\) pattern.
These patterns are graphs which are mapped to each other via graph morphisms \(l,r\) from or into a third graph \(I\), such that \(L, R, I\) constitute either a span or a co-span, respectively~\cite{ehrig2006fundamentals,ehrig2009alternative,mantz2015coevolving}.
In this paper we use the co-span version, which facilitates the manipulation of model elements without the two phases of delete-then-add.
This same construction will be reused for the next two alternatives.

In order to apply a rule to a source graph \(S\), first a match of the left pattern must be found in \(S\), i.e., a graph homomorphism \(m:L \rightarrow S\).
Then, using a pushout construction, followed by a pullback complement construction, will create a target graph \(T\).
Details of application conditions and theoretical results on graph transformations can be found in~\cite{ehrig2006fundamentals}.
We omit in our sample rules the details introduced by graph \(I\).

In typed transformation rules, the calculation of matches needs to fulfil a typing condition: In the construction in Figure~\ref{fig:two-level-rule-formal}, all the triangles must be commutative.
In other words, since both the rules and the models are typed by the same type graph, the rules are defined at the type graph level and are applied to typed graphs.

\subsubsection{Multilevel Rules}
\begin{figure}
	\centering
	\includegraphics[scale=.9]{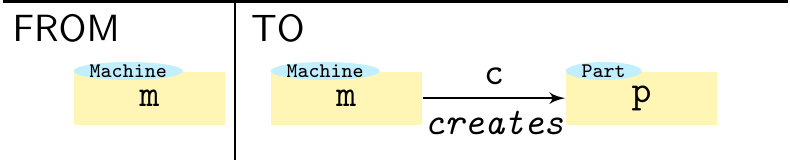}
	\caption{Abstract behaviour for generator-like machines}
	\label{fig:pls-rule-create-part-multilevel}
\end{figure}

Multilevel rules, also known as deep rules (see, e.g.,~\cite{deLara2015}), enable us to specify a common behaviour at upper levels of abstraction.
The creation of parts in our PLS example may be specified using multilevel rules by a single rule, operating on the more abstract types \elementname{Machine} and \elementname{Part} as displayed in Figure~\ref{fig:pls-rule-create-part-multilevel}.
Thus, this rule can be applied to \elementname{GenHandle} and \elementname{GenHead} in the hammer plant (Figure~\ref{fig:pls-multilevel-hierarchy}b) as well as to \elementname{GenLeg} and \elementname{GenSeat} generators in the stool plant (Figure~\ref{fig:pls-multilevel-hierarchy}c).

\begin{wrapfigure}[10]{r}[0mm]{.5\textwidth}
	\centering
	\vspace{-6mm}
	\includegraphics[scale=.85]{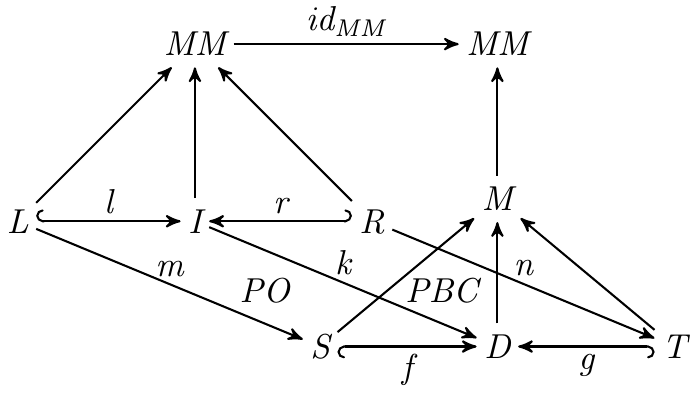}
	\caption{Multilevel MT rule}
	\label{fig:multilevel-rule-formal}
\end{wrapfigure}

While multilevel model transformations solve some problems, it introduces a new issue related to case distinction.
That is, the approach works fine in cases where the model on which the rules are applied --- usually a running instance --- contains the structure that is required by the rule, and, when all types required by the rule exist in the same metalevel.
If not, the rules will only be able to express behaviour in a generic way, but they will not be precise enough.
In other words, the rules become too generic and imprecise: all machines will create parts (even machines which should not be creating parts), all parts can be created directly (even those which should be derived from others), and any machine can create any part.
For example, \elementname{GenHead} may create \elementname{Hammer}, and \elementname{GenHandle} may create a \elementname{Head}.

The general structure of multilevel transformation rules is shown in Figure~\ref{fig:multilevel-rule-formal}.
This could be considered as a method to relax the strictness of two-level model transformations through multilevel model transformations (see, e.g.,~\cite{atkinson2012towards,atkinson2015enhancing}).
To achieve flexibility, the rules could be defined over a type graph somewhere at a higher level in the hierarchy and applied to running instances at the bottom of the hierarchy (see Figure~\ref{fig:multilevel-rule-formal}).
Types will be resolved by composing typing graph homomorphisms from the model on which the rule is applied and upwards to the level where of the type graph.

\subsubsection{Multilevel Coupled Model Transformation Rules}
\begin{figure}
	\centering
	\includegraphics[scale=.75]{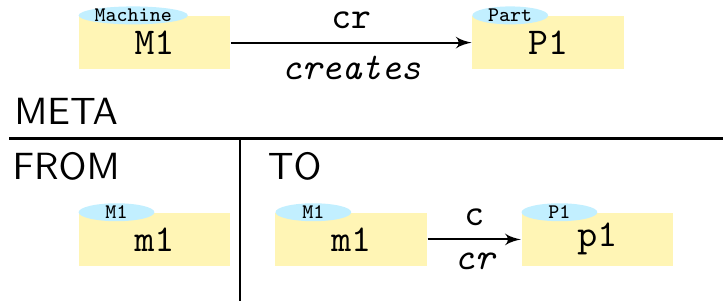}
	\caption{Rule \textit{CreatePart}: a machine creates a part}
	\label{fig:pls-rule-create-part}
\end{figure}

We propose Multilevel Coupled Model Transformation (MCMT) rules, as a means to overcome the issues of the two aforementioned approaches.
Using MCMTs, the rule \textit{CreatePart} can be specified as shown in Figure~\ref{fig:pls-rule-create-part}.
We introduce a new component into the rules, namely the META block.
This new part of the rule allows us to locate types in any level of the hierarchy that can then be used in the FROM and TO blocks.
Moreover, the real expressive power from this new block comes from the fact that we can define a whole pattern that these types must satisfy in order for the rule to be applied.
This feature allows us to create case distinctions in the rules easily, allowing for a finer tuning of the rules that prevents the aforementioned side effects.
This version of \textit{CreatePart} can be applied on any PLS typed by \emph{Generic Plant}, since the variable \elementname{P1} could match any of the parts --- \elementname{Head}, \elementname{Handle}, \elementname{Seat} or \elementname{Leg} --- and the variable \elementname{M1} could match any of the \textit{creator} machines --- \elementname{GenHead}, \elementname{GenHandle}, \elementname{GenSeat} or \elementname{GenLeg}.
However, the important difference is that this rule can only match the generators of parts, since we require \elementname{M1} in the META block to have a \elementname{creates} relation to the \elementname{P1}.

We find a match of this rule when we find an element which, when coupled together with its type, fits an instance of \elementname{M1} that has a relation of type \elementname{creates} to an instance of \elementname{P1}.
For example, \elementname{GenHandle} in Figure~\ref{fig:pls-multilevel-hierarchy}b, fits \elementname{M1}, since \elementname{GenHandle} has a \elementname{creates} relation to \elementname{Handle}.
Hence, \elementname{m1} can be matched to \elementname{ghandle} when applying the rule, in order to create a new part (\elementname{p1}), which would be an instance of \elementname{Handle}.

The usage of this new block allows for a finer tuning that other techniques, like inheritance.
For example, the possibility of referring to the \elementname{creates} relation between \elementname{Machine} and \elementname{Part} when more abstraction is required, while still being able to instantiate it only in some cases (like \elementname{GenHead} and \elementname{Head}) is a level of granularity cannot be achieved by turning all instances of \elementname{Machine} and \elementname{Part} into subclasses.
This would render impossible to retain the abstraction ``machines can create parts'' while specifying that ``generators of heads generate heads'' and that ``assemblers do not create any part''.
Therefore, this alternative would create undesirable side effects.

\begin{wrapfigure}[21]{r}[0mm]{.4\textwidth}
	\centering
	\includegraphics[width=.4\textwidth]{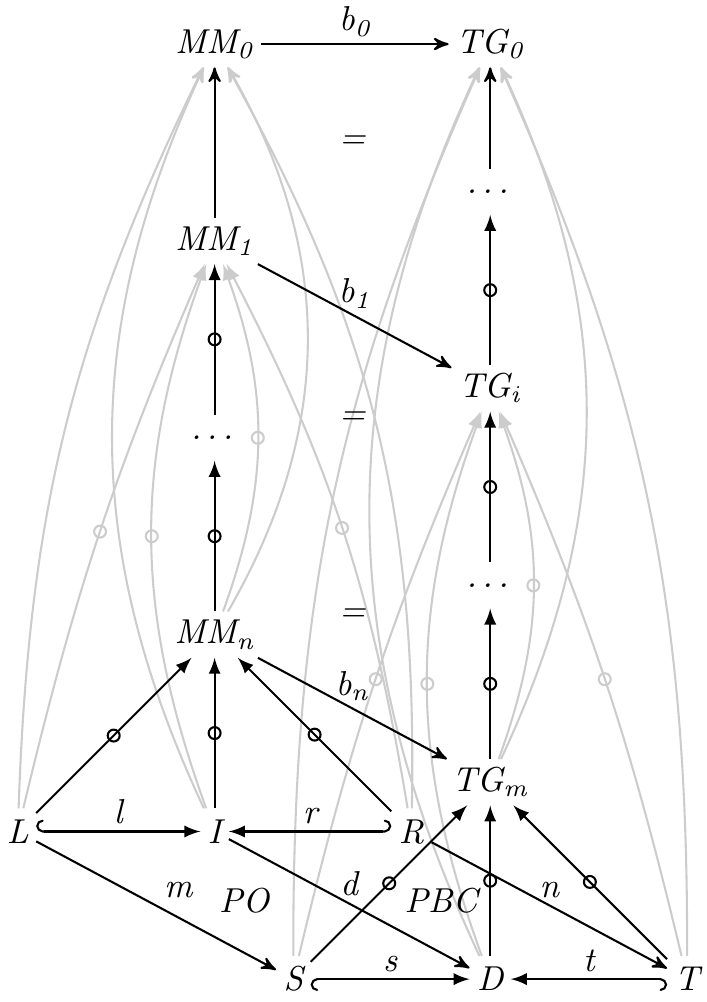}
	\caption{Formal diagram of an MCMT}
	\label{fig:multilevel-coupled-rule-formal}
\end{wrapfigure}

The META section may include several metamodel levels. 
The general structure of an MCMT is displayed in Figure~\ref{fig:multilevel-coupled-rule-formal}.
The figure can be visualized as two flat trees, each of them defined by typing chains and connected to each other by matching morphisms.

The tree on the left contains the pattern that the user defines in the rule.
It consists of the left and right parts of the rule (FROM and TO, respectively), represented as \emph{L} and \emph{R} in the diagram, and the interface \emph{I} that contains the union of both \emph{L} and \emph{R}, hence the inclusion morphisms\footnote{In all our examples in this paper, the interface \emph{I} contains the left and right graphs since the morphisms \emph{l} and \emph{r} are monomorphisms, but this might not be the general case.}.
These three graphs are typed by elements in the same typing chain, which is represented as a sequence of metamodels \emph{MM\textsubscript{x}} that ends with the root of the hierarchy tree \emph{MM\textsubscript{0}} (Ecore in our case).
The multilevel typing arrows from $L,I,R$ to the typing chain $[\textit{MM}_n, \textit{MM}_{n-1}, \dots, \textit{MM}_1, \textit{MM}_0]$ (see Section~\ref{subsec:multilevel-typing}) should be compatible with typing (see Section~\ref{subsec:compatibility-typing}).
That is, both triangles at the bottom must be commutative.

The tree on the right side represents the actual path in the type chain on which the rule is applied.
Before applying the rule, we only have the graph \emph{S} and its multilevel type to the typing chain $[\textit{TG}_m, \textit{TG}_{m-1}, \dots, \textit{TG}_1,\textit{TG}_0]$.
In order for the rule to be applied, it is required to find matches of all metamodel graphs \emph{MM\textsubscript{x}} into the actual typing chain graphs \emph{TG\textsubscript{y}}.
These matches do not require to be \textit{parallel} in the sense that the difference of levels between the two sources of any two matching morphisms is not required to be equal to the difference of levels between the targets of those two morphisms.
This is due to the flexibility in the specification of the number of levels that separate two metamodel graphs in the pattern.
In terms of our example, if we add more intermediate levels to the PLS hierarchy, and consequently the depth becomes bigger, the defined rules would still be applicable.
Moreover, when the rules are defined with a flexible depth, they would fit or match different branches of the hierarchy regardless their depth.
That is, the graph representing the pattern could be matched in several different ways to the same hierarchy, hence providing the flexibility that we require.

The match \emph{b\textsubscript{0}} is trivial in the implementation since both \emph{MM\textsubscript{0}} and \emph{TG\textsubscript{0}} are Ecore.
If all the metamodel graphs can be matched into the type graphs, such that every resulting square commutes, we proceed to find a match (a graph homomorphism $m$) of the pattern graph \emph{L} into the instance \emph{S}.
This match must be type compatible, i.e., the types of elements in $L$ which are matched to elements in $S$ must be matched by one of the bindings $b_x$ (see Section~\ref{subsec:compatibility-typing}).
If this match \emph{m} is successful, i.e., a type compatible graph homomorphism from $L$ to $S$ is found, we construct the intermediate instance \emph{D} by pushout, and then proceed to generate the target instance \emph{T} by pullback complement.
The necessary formal constructions are introduced in Section~\ref{sec:chain-category}.

To summarise, while existing approaches which employ reusable model transformations for the definition of behavioural models focus on traditional two-level modelling hierarchies and their affiliated two-level model transformations (see~\cite{kusel2015reuse,kusel2013reality} for a survey), multilevel model transformations~\cite{atkinson2015enhancing,atkinson2012towards} are relatively new and are not yet proven suitable for reuse and definition of model's behaviour.
Our approach to multilevel model transformation build on top of these original approaches to combine reusability with flexibility in the number of modelling levels.
In other words, in the time of defining the rules, the height/depth of the modelling hierarchy (on which we intend to apply the rules) is not important since no matter how deep the hierarchy gets, the rules will still work.
It is also important to notice that, so far, we have not identified a scenario in which MCMTs have more expressive power than two-level MTs.
That is, one MCMT conveys the same information as an arbitrarily large set of two-level MTs with commonalities, but in a much more concise, reusable and maintainable way, as remarked in Section~\ref{sec:conclusions-and-future-work}.

\subsection{MCMTs for PLS}
\label{subsec:mcmt-pls}

Using the PLS example, we have already explained the advantages of using MCMTs with respect to reusability and shortly presented a comparison to two-level and multilevel model transformation rules.
Here we will show a few MCMTs which together define the behaviour of any plant which directly or indirectly conforms to \elementname{generic\_plant}.

The first rule in the PLS example is \textit{CreatePart}, already shown in Figure~\ref{fig:pls-rule-create-part}, which is used to create parts by the corresponding generator machines.
Note that, as mentioned above, this rule is not restricted to instances of \elementname{hammer\_plant} and it could also be applied to any instance --- direct or indirect --- of \elementname{generic\_plant}.

\begin{figure}[tb]
	\centering
	\includegraphics[width=\linewidth]{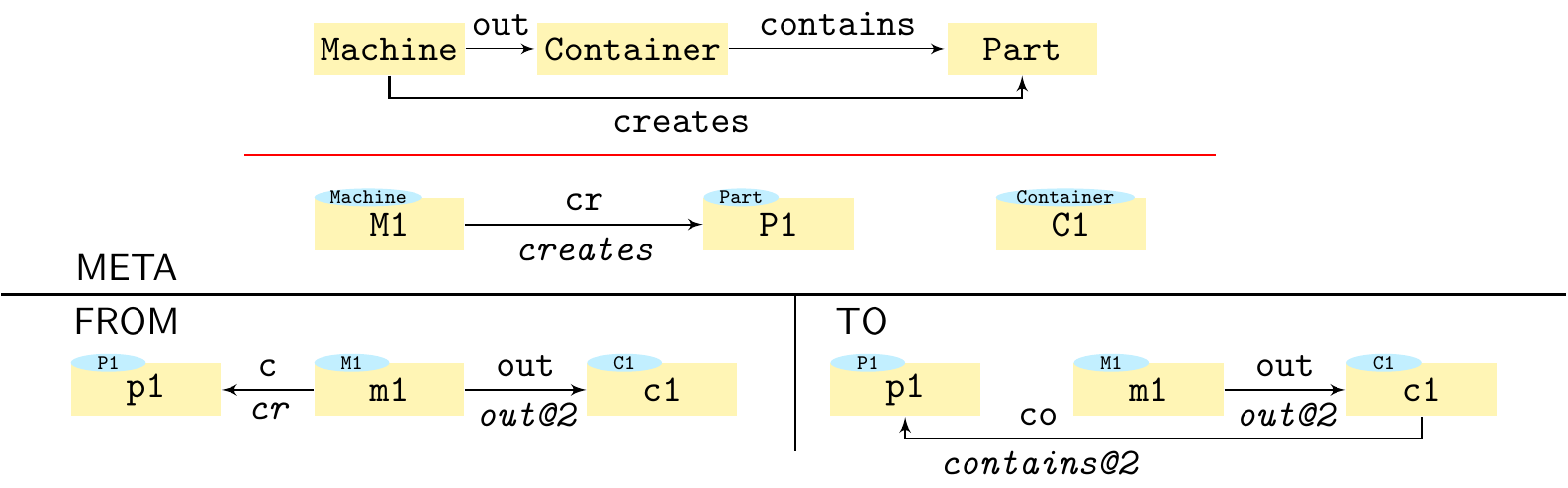}
	\caption{Rule \textit{SendPartOut}: a part leaves the machine that created it}
	\label{fig:pls-rule-send-part-out}
\end{figure}

Another rule used in this example is \textit{SendPartOut}, shown in Figure~\ref{fig:pls-rule-send-part-out}, used for moving a created part from its generator into the output container of this machine.
This rule displays a richer META block than \textit{CreatePart}, in which we need to identify elements from two different models, separated with a red line, as in the examples in Section~\ref{sec:hierarchies}.
Such information was not displayed explicitly in \textit{CreatePart} since it could be automatically inferred and for the sake of simplicity.
At the top level, we mirror part of \elementname{generic\_plant}, defining elements like \elementname{out} and \elementname{contains}, that are used directly as types in the FROM and TO blocks, via potency.
These elements are defined as constants, meaning that the name of the pattern element must match an element with the same name in the typing chain.
The use of constants allows us to be more restrictive when matching, and significantly reduces the amount of matches that we obtain (see Section~\ref{subsec:matching-algorithm}). 

\begin{figure}
	\centering
	\includegraphics[width=.8\linewidth]{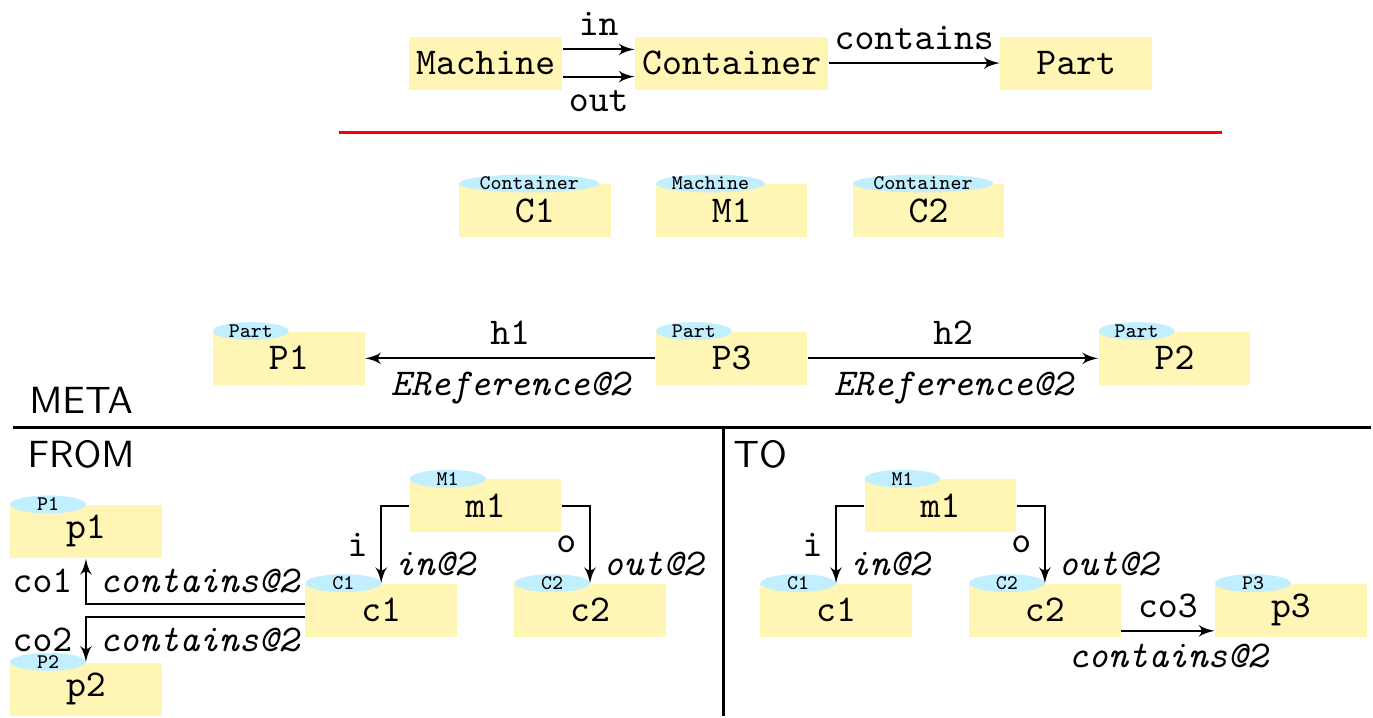}
	\caption{Rule \textit{Assemble}: a machine assembles two parts into a new one}
	\label{fig:pls-rule-assemble}
\end{figure}

Once the component parts have reached the right containers, it is possible to combine them to create final products --- hammers or stools in our hierarchy.
The \textit{Assemble} rule does exactly that: it assembles two parts into a different part (see Figure~\ref{fig:pls-rule-assemble}).
It requires the resulting part \elementname{p3} to be a part that can consist of, or be built from, parts \elementname{p1} and \elementname{p2}.
One particularity of this rule is that the META block defines several instances of the same type, but different from each other.
In such a way, we ensure that, for example, \elementname{P1} is different from \elementname{P2}, but both are instances of \elementname{Part}.

The semantics of matching makes this rule applicable for a variable number of parts which are assembled into one part.
To apply this rule we use the cardinality information of relations \elementname{h1} and \elementname{h2}, as long as the upper bound is different from \elementname{*}.
The cardinalities of \elementname{h1} and \elementname{h2} in the corresponding model will indicate the number of instances of \elementname{P1} and \elementname{P2} that should appear in the FROM block.
This information is then used to replicate the instances \elementname{p1} and \elementname{p2} accordingly, together with all their related references (\elementname{co1} and \elementname{co2}, in this case).
That is, this rule fits both the \elementname{Gluer} and the \elementname{Assembler} functionalities of the two sample plants (for hammers and stools).
For example, this process gives us the semantics that, in order to build an instance of \elementname{Stool}, we require three instances of \elementname{Leg} and one instance of \elementname{Seat}.
In case that the cardinalities allow for several values, i.e., the minimum and maximum values are different, one copy of the rule will be generated for each possible value.

\begin{figure}
	\centering
	\includegraphics[width=\linewidth]{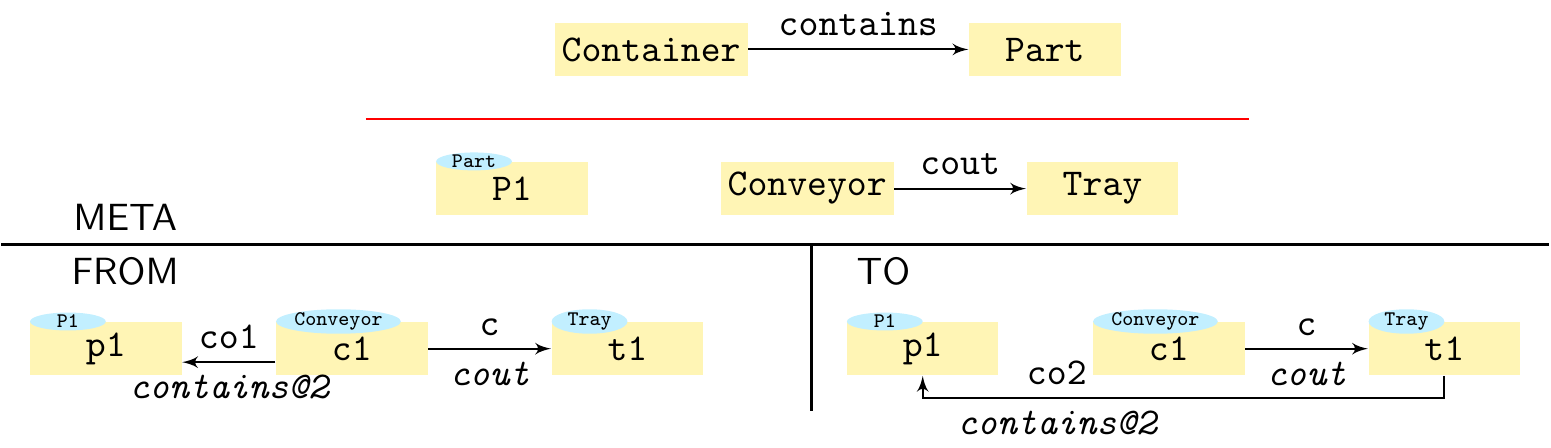}
	\caption{Rule \textit{TransferPart}: a part is transferred from a conveyor into a tray}
	\label{fig:pls-rule-transfer-part}
\end{figure}

Finally, the last rule is used to specify the transfer of the assembled part from the assembler's output conveyor into a tray.
This behaviour is achieved by applying the rule \textit{TransferPart} displayed in Figure~\ref{fig:pls-rule-transfer-part}.

\section{The Category of Graph Chains and Graph Chain Morphisms}
\label{sec:chain-category}

We use model transformations to define the behaviour of our MLM hierarchies.
Since every branch in our MLM hierarchies is represented by a graph chain, the application of our model transformation rules will rely on pushout and pullback complement constructions in the category \cat{Chain}.
By constructing this category for multilevel hierarchies and MCMTs, we are able build upon the already existing co-span approach to graph transformations~\cite{ehrig2009alternative}.
We introduce appropriate definitions for the category \cat{Chain} in this section.

This formalisation is required since the semantics of model transformations needs to be adapted to our multilevel setting, keeping flexibility of rules in mind.
The formalism can be later used as a reference for the expected behaviour of MCMTs, regardless of the mechanism used to implement them (e.g. the proliferation process presented in section~\ref{subsec:proliferation}).
Detailed descriptions and proofs of the constructions are left out as a technical report~\cite{macias2017chains}.

\subsection{Multilevel typing}
\label{subsec:multilevel-typing}

MCMTs are defined as graphs $L,I,R$ which are located at the bottom of a typing chain representing the metamodels in the rule hierarchy.
Moreover, the instance graph $S$ on which we apply the MCMTs is also located at the bottom of the typing chain.
We call the relation between these graphs and their typing chains \textit{multilevel typing}.
In fact, the relation between each graph $G_i$ in a typing chain $\tc{G}$ and the rest of the chain $[G_{i-1} ... G_0]$ above the graph is a multilevel typing.
Further, applying an MCMT will produce the intermediate graph $D$ and the target graph $T$ which have to have a multilevel typing relation to the target typing chain.
To be able to use the well-established constructions of pushout and pullback complement in the lowest level graphs $L$, $I$, $R$, $S$, $D$, and $T$, and indeed get the multilevel typing of the constructed graphs, we will define some restrictions on typing chains.
First of all, we will generalise the concept of typing chain and formally define the concept ``graph chain''.
Then, we will define graph chain morphisms and multilevel typing of a graph over a graph chain.

\begin{definition}[Graph chain]
\label{def:graph-chain}
A graph chain \(\chainname{G} := \chain{G}{n}{\tau^G}\) is given by a natural number $n$, a sequence \(\overline{G}=[\graphname{G}{n}{}, \graphname{G}{n-1}, \dots, \graphname{G}{1}{}, \graphname{G}{0}{}]\) of graphs of length \(n+1\)
together with a family \(\tau^G=(\tau^G_{j,i}:\graphname{G}{j}\partialmap\graphname{G}{i}\mid 0\le\level{i}<\level{j}\le\level{n})\) of partial graph homomorphisms where
\begin{itemize}
	\item each partial graph homomorphism \(\tau^G_{j,i}:\graphname{G}{j}\partialmap\graphname{G}{i}\) is given by a subgraph \(\domain{\tau^G_{j,i}}\sqsubseteq\graphname{G}{j}\), called the domain of definition of \(\tau^G_{j,i}\), and a total graph homomorphism \(\tau^G_{j,i}:D(\tau^G_{j,i}) \xrightarrow{} G_i \),
	\item all the morphisms \(\tau^G_{j,0}:\graphname{G}{j}\to\graphname{G}{n}\) with \(1\le\level{j}\le\level{n}\) to the top level graph are total, i.e., \(D(\tau^G_{j,0})=G_j\), and
	\item for all \(0\le\level{i}<\level{j}<\level{k}\leq n\) the uniqueness condition \(\tau^G_{k,j};\tau^G_{j,i}\preceq\tau^G_{k,i}\) is satisfied, i.e., \(D(\tau^G_{k,j};\tau^G_{j,i})\sqsubseteq D(\tau^G_{k,i})\) and, moreover, \(\tau^G_{k,j};\tau^G_{j,i}\) and \(\tau^G_{k,i}\) coincide on \(D(\tau^G_{k,j};\tau^G_{j,i})\). The composition \(\tau^G_{k,j};\tau^G_{j,i}\) is defined by pullback (inverse image) of total graph homomorphisms as follows:
	
	\begin{center}
			$\xymatrix@R=4mm{
				&& D(\tau^G_{k,j};\tau^G_{j,i}) \ar@{_{(}->}[dl]_-{\sqsubseteq} \ar[dr] \ar@/^12mm/[ddrr]^{\tau^G_{k,i}} \\
				&   D(\tau^G_{k,j}) \ar@{_{(}->}[dl]_-{\sqsubseteq} \ar[dr]^{\tau^G_{k,j}}
				& PB
				& D(\tau^G_{j,i}) \ar@{_{(}->}[dl]_-{\sqsubseteq} \ar[dr]^{\tau^G_{j,i}} \\
				G_k && G_j && G_i
			}$\\
	\end{center}
	
	That is, \(D(\tau^G_{k,j};\tau^G_{j,i}) = (\tau^G_{k,j})^{-1} (D(\tau^G_{j,i}) ) \) thus we have \(D(\tau^G_{k,j};\tau^G_{j,i}) = D(\tau^G_{k,j})\) if \(\tau^G_{j,i}\) is total, i.e., \(G_j = D(\tau^G_{j,i})\).
\end{itemize}

\end{definition}

Applying an MCMT rule would now require finding a match of the rule, which has three characteristics: (i) a match homomorphism from $L$ to $S$, (ii) a match of the graph chain $\cal{MM}$ into the graph chain $\cal{TG}$, and, (iii) both these matches need to be \emph{compatible} with respect to multilevel typing.
To formalise the intended flexibility of matching MCMT rules, and in order to define matches between graph chains, we define \emph{graph chain morphisms}.

\begin{definition}[Graph chain morphism]
\label{def:graph-chain-morphism}
A morphism \( (\phi,f) : \chainname{G} \to \chainname{H}\) between two graph chains $\chainname{G}=\chain{G}{n}{\tau^G}$ and $\chainname{H}=\chain{H}{m}{\tau^H}$ with $n\leq m$ is given by
\begin{itemize}
	\item a function \(f:[n] \to [m] \), where \( [n] \)  \( \in \{ 0,1, 2,\ldots,n\} \), such that $f(0) = 0$ and $i<j$ implies $f(i) < f(j)$ for all $i,j\in[n]$, and
	\item a family of total graph homomorphisms $\phi = (\phi_i : G_i \to H_{f(i)} \mid i \in [n] )$ such that $\tau^G_{j,i};\phi_i = \phi_j;\tau^H_{f(j),f(i)}$ for all $0 \leq i < j \leq n$, i.e., due to the definition of composition of partial morphisms (cf. Definition~\ref{def:graph-chain}), we assume for any $0 \leq i < j \leq n$ the following commutative diagram of total graph homomorphisms
\begin{center}
	$\xymatrix@R=5mm{
		G_i \ar[rr]^{\phi_i} && H_{f(i)}\\
		D(\tau^G_{j,i}) \ar@{_{(}->}[d]_-{\sqsubseteq} \ar@{.>}[rr]^{\phi_{j|i}} \ar[u] \ar@{}[drr]|{PB}
		&& D(\tau^H_{f(j),f(i)}) \ar@{_{(}->}[d]_-{\sqsubseteq} \ar[u] \\
		G_j \ar@/^12mm/[uu]^{\tau^G_{j,i}}|{\circ} \ar[rr]^{\phi_{j}}
		&& H_{f(j)} \ar@/_12mm/[uu]_{\tau^H_{f(j),f(i)}}|{\circ}
	}$\\
\end{center}
\end{itemize}
\end{definition}

A match of the rule graph chain \(\chainname{MM} := \chain{MM}{n}{\tau^{MM}}\) to the target graph chain \(\chainname{TG} := \chain{TG}{m}{\tau^{TG}}\) is now defined as a graph chain morphism \( (b,f) : \chainname{MM} \to \chainname{TG}\).

Note that since the composition of partial morphisms is based on a pullback construction, the condition $\tau_{j,i}^{MM};b_i = b_j;\tau_{f(j),f(i)}^{TG}$, for all $0 \leq i < j \leq n$, means that we are not only requiring that typing is preserved, but that typing is also reflected, since $b_j$ and $b_i$ are total morphisms.
That is, we have:
\begin{equation}\label{eq:preserve_reflect}
	D(\tau_{j,i}^{MM}) = b_j^{-1}(D(\tau_{f(j),f(i)}^{TG}))
\end{equation}

\begin{remark}\label{rem:po}
Identity graph chain morphisms and composition of graph chain morphisms are straightforward to define and to be shown to satisfy the category axioms, and thus we obtain the category $\cat{Chain}$ of graph chains and graph chain morphisms.
\end{remark}

As mentioned, all the graphs $L,I,R,S,D,T$ in Figure~\ref{fig:multilevel-coupled-rule-formal} are multilevel-typed over either the graph chain $\cal{MM}$ or $\cal{TG}$.
To explain the multilevel typing relation we consider one of these graphs, $S$, and one of the graph chains, $\cal{TG}$, however, the discussion can be generalised to any of the other graphs and graph chains.

\begin{figure}
	\begin{minipage}[b]{.44\linewidth}
	\centering
	\begin{tikzpicture}[on grid,node distance=15mm]
			
		\tikzset{element/.style={inner sep=2pt,minimum height=1.3em,font=\itshape}}
		\tikzset{label/.style={auto,midway,above,font=\itshape,inner sep=2pt}}
		\tikzset{map/.style={->,>=stealth',semithick}}
		\tikzset{partmap/.style={map,postaction={decorate},decoration={markings,mark=at position 0.5 with {\draw circle [radius=.4ex];}}}}
		
		\node[element] (s)							{S};
		\node[element] (tg0)	[right=25mm of s]	{\typegraph{0}};
		\node[element] (tg1)	[below of=tg0]		{\typegraph{1}};
		\node[element] (d1)		[below of=tg1]		{\vdots};
		\node[element] (tgi)	[below of=d1]		{\typegraph{i}};
		\node[element] (d2)		[below of=tgi]		{\vdots};
		\node[element] (tgj)	[below of=d2]		{\typegraph{j}};
		\node[element] (d3)		[below of=tgj]		{\vdots};
		\node[element] (tgk)	[below of=d3]		{\typegraph{k}};
		\node[element] (d4)		[below of=tgk]		{\vdots};
		\node[element] (tgm)	[below of=d4]		{\typegraph{m}};
		
		\node[element] (eq1)	[below left=10mm of tg0]	{$\preceq$};
		\node[element] (eq2)	[below of=eq1]				{$\preceq$};
		\node[element] (eq3)	[below=25mm of eq2]			{$\preceq$};
		\node[element] (eq4)	[below=30mm of eq3]			{$\preceq$};
		\node[element] (eq5)	[below=35mm of eq4]			{$\preceq$};
		
		\draw[map]		(s)			to					node [label]		(s0)	{\chainmorph{0}}	(tg0);
		\draw[partmap]	(s)			to[bend right=5] 	node [label]		(s1)	{\chainmorph[S]{1}}	(tg1);
		\draw[partmap]	(s)			to[bend right=5] 	node [label,right]	(si)	{\chainmorph[S]{i}}	(tgi);
		\draw[partmap]	(s)			to[bend right=10] 	node [label,right]	(sj)	{\chainmorph[S]{j}}	(tgj);
		\draw[partmap]	(s)			to[bend right=15] 	node [label,right]	(sk)	{\chainmorph[S]{k}}	(tgk);
		\draw[partmap]	(s)			to[bend right=20] 	node [label,right]	(sn)	{\chainmorph[S]{m}}	(tgm);
		
		\draw[map]		(tg1)		to					node [label,right]	(t10)	{\typemorph[TG]{1}{0}}	(tg0);
		\draw[partmap]	(d1)		to 					node [label]		(td1)	{}					(tg1);
		\draw[partmap]	(tgi)		to					node [label]		(tid)	{}					(d1);
		\draw[partmap]	(d2)		to 					node [label]		(tdi)	{}					(tgi);
		\draw[partmap]	(tgj)		to					node [label]		(tjd)	{}					(d2);
		\draw[partmap]	(d3)		to 					node [label]		(tdj)	{}					(tgj);
		\draw[partmap]	(tgk)		to					node [label]		(tkd)	{}					(d3);
		\draw[partmap]	(d4)		to 					node [label]		(tdk)	{}					(tgk);
		\draw[partmap]	(tgm)		to					node [label]		(tnd)	{}					(d4);
		
		\draw[map]		(tgi.east)	to[bend right=25]	node [label,right]	(ti0)	{\typemorph[TG]{i}{0}}	(tg0.east);
		\draw[partmap]	(tgj.east)	to[bend right=15]	node [label,right]	(tji)	{\typemorph[TG]{j}{i}}	(tgi.east);
		\draw[partmap]	(tgk.east)	to[bend right=15]	node [label,right]	(tkj)	{\typemorph[TG]{k}{j}}	(tgj.east);
		\draw[partmap]	(tgk.east)	to[bend right=45]	node [label,right]	(tki)	{\typemorph[TG]{k}{i}}	(tgi.east);
		
		\end{tikzpicture}
		\subcaption{Multilevel typing of the graph $S$ over $\cal{TG}$}
		\label{fig:mlt_s_tg}
	\end{minipage}
	\quad
	\begin{minipage}[b]{.44\linewidth}
		\begin{center}
			$\xymatrix@R=11mm@C=4mm{
				S_0 = S \ar[rr]^{\sigma^S_0}
				&& TG_0\\
				S_1 = D(\sigma^S_1) 
				  \ar[rr]^{\sigma^S_1} 
				  \ar@/^10mm/@{_{(}->}[u]
				&& TG_1 \ar[u]\\
				\vdots && \vdots  \ar[u]|{\circ}\\
				S_i = D(\sigma^S_i) 
				  \ar[rr]^{\sigma^S_i} 
				  \ar@/^18mm/@{_{(}->}[uuu]
				&& TG_i \ar@/_10mm/[uu]_(0,8){\tau^{TG}_{i,k}}|{\circ} \ar@/_15mm/[uuu]_{\tau^{TG}_{i,0}}  \ar[u]|{\circ}\\
				\vdots && \vdots  \ar[u]|{\circ}\\				
				S_j = D(\sigma^S_j) 
				  \ar[rr]^{\sigma^S_j} 
				  \ar@/^18mm/@{_{(}->}[uuuuu]
				  \ar@/^5mm/@{.>}[uu]^{\tau^S_{j,i}}|{\circ}
				&& TG_j  \ar[u]|{\circ}
				  \ar@/_10mm/[uu]_{\tau^{TG}_{j,i}}|{\circ}\\
				\vdots && \vdots  \ar[u]|{\circ} \\
				S_m = D(\sigma^S_m) \ar[rr]^{\sigma^S_m} \ar@/^18mm/@{_{(}->}[uuuuuuu]
				&& TG_m \ar@/_10mm/[uu]_{\tau^{TG}_{m,j}}|{\circ}   \ar[u]|{\circ}\\
			}$\\
		\end{center}
		\subcaption{A graph and its multilevel typing represented as an inclusion chain and a chain graph morphism}
		\label{fig:mlt_to_chaingraph}
	\end{minipage}%
	\caption{Refactoring multilevel typing to chain graph and chain graph morphism}
	\label{fig:image2}
\end{figure}
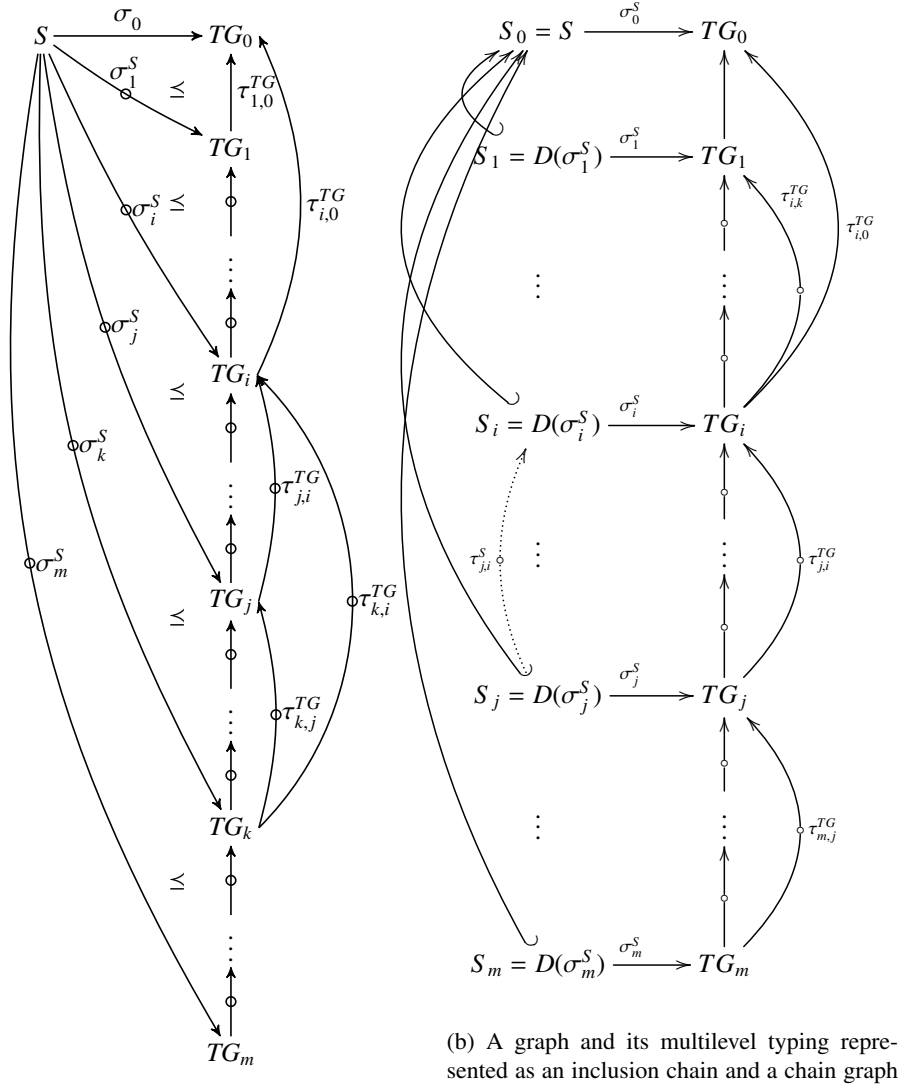

The multilevel typing of $S$ over $\cal{TG}$ is given by a family of partial typing morphisms $\sigma^S = {(\sigma_i^S : S \partialmap TG_i \mid 0 \leq i \leq m)}$ where $\sigma_0^S : S \to TG_0$ is total (see Figure~\ref{fig:mlt_s_tg}).
This multilevel typing must be compatible with the ``internal'' typing morphisms $\tau^{TG}$.
Analogous to the uniqueness condition for graph chains we require that
\begin{equation}\label{eq:u2}
	\sigma^S_j; \tau_{j,i}^{TG} \preceq \sigma_i^S \,\,\,\, \textit{for all} \,\,\,\, 0 \leq i \leq j \leq m
\end{equation}

The family $\sigma^S$ of partial typing morphisms defines implicitly a sequence $\bar{S} = [S_m, S_{m-1}, \dots, S_1, S_0]$ of subgraphs of $S$ with $S_0:=S$ and $S_i = D(\sigma^S_i)$, for all $0\leq i \leq m$ (see Figure~\ref{fig:mlt_to_chaingraph}).
Due to the composition of partial morphisms, the condition~(\ref{eq:u2}) is equivalent to the condition
\begin{equation}\label{eq:u3}
	(\chainmorph[S]{j})^{-1}(D(\typemorph[TG]{j}{i})) \sqsubseteq S_i := D(\chainmorph[S]{i})
\end{equation}
That is, if we assign to an item $e$ in $S$ a type $\chainmorph[S]{j}(e)$ in $\typegraph{j}$ and this item, in turn, has a transitive type $\typemorph[TG]{j}{i}(e)$ in $\typegraph{i}$ then $\chainmorph[S]{i}$ has to assign this transitive type to $e$.
It is adequate to require a stronger condition that if both $\chainmorph[S]{j}$ and $\chainmorph[S]{i}$ assign a type to $e$ then $\chainmorph[S]{i}(e)$ has to be a transitive type of $\chainmorph[S]{j}(e)$.
Note that $	(\chainmorph[S]{j})^{-1}(D(\typemorph[TG]{j}{i})) \sqsubseteq S_j = D(\chainmorph[S]{j})$ by definition of pre-images.
In such a way, we require that
\begin{equation}\label{eq:u4}
	(\chainmorph[S]{j})^{-1}(D(\typemorph[TG]{j}{i})) = S_j \cap S_i = D(\chainmorph[S]{j}) \cap D(\chainmorph[S]{i})  \,\,\,\, \textit{for all} \,\,\,\, 0 \leq i \leq j \leq m
\end{equation}
That is, we require the following diagram to commute for all $0 \leq i \leq j \leq m$

\begin{center}
			$\xymatrix@R=5mm{
				S_i = D(\sigma^S_i) \ar[rr]^{\sigma^S_i}
				&& TG_i\\
				S_{j} \cap S_{i} \ar@{_{(}->}[d]_-{\sqsubseteq} \ar@{.>}[rr] \ar@{^{(}->}[u]^-{\sqsubseteq} \ar@{}[drr]|{PB}
				&& D(\tau^{TG}_{j,i}) \ar@{_{(}->}[d]_-{\sqsubseteq} \ar[u]_{\tau^{TG}_{j,i}} \\
				S_j \ar[rr]^{\sigma^S_j}
				&& TG_j \ar@/_18mm/[uu]_{\tau^{TG}_{j,i}}|{\circ}\\
			}$\\
	\end{center}

By using the stronger yet adequate condition~(\ref{eq:u4}) we can now:
\begin{itemize}
	\item Transform any graph $S$ into an abstract multilevel typing structure independent of any other typing chain, by simply choosing a sequence of arbitrary subgraphs of $S$ (see Corollary~\ref{cor:refactoring}).
	\item Express an actual multilevel typing over a given graph chain by means of chain morphisms.
\end{itemize}

\begin{corollary}[Refactoring]\label{cor:refactoring}
For any graph $S$ we can extend any sequence $\tc{S} = [S_m, S_{m-1}, \dots, S_1, S_0]$ of subgraphs of $S$, with $S_0 = S$, to a graph chain $\cal{S}$ $ = (\tc{S}, m, \typemorph[S]{}{})$ where for all $0 \leq i \leq j \leq m$, $\typemorph[S]{j}{i} : S_j \partialmap S_i$ (also called a partial inclusion morphism) is given by $D(\typemorph[S]{j}{i}) := S_j \cap S_i$ and the span ${S_j \hookleftarrow D(\typemorph[S]{j}{i}) := S_j \cap S_i \hookrightarrow S_i}$ of inclusions.
We call the graph chain ${\cal{S} = (\tc{S}, m, \typemorph[S]{}{})}$ also an \textit{inclusion chain}.
\end{corollary}

To summarise, for any of the pairs $(L, \chainmorph[L]{})$, $(I, \chainmorph[I]{})$, $(R, \chainmorph[R]{})$ and $(S, \chainmorph[S]{})$, of a graph and a multilevel typing, we require that condition~(\ref{eq:u4}) is satisfied,
which ensures that we get by Corollary~\ref{cor:refactoring} four corresponding inclusion chains 
$\cal{L} $$ = (\tc{L}, n,\typemorph[L]{}{})$,  
$\cal{I} $$ = (\tc{I}, n,\typemorph[I]{}{})$,  
$\cal{R} $$ = (\tc{R}, n,\typemorph[R]{}{})$ and  
$\cal{S} $$ = (\tc{S}, m,\typemorph[S]{}{})$, respectively, 
together with chain morphisms 
$(\chainmorph[L]{},id_{[n]}) : $ $\cal{L}$ $\to$ $\cal{MM}$, 
$(\chainmorph[I]{},id_{[n]}): $ $\cal{I}$ $\to$ $\cal{MM}$, 
$(\chainmorph[R]{},id_{[n]}): $ $\cal{R}$ $\to$ $\cal{MM}$, and
$(\chainmorph[S]{},id_{[m]}): $ $\cal{S}$ $\to$ $\cal{TG}$.
Note that the $\sigma$'s in the chain morphisms are the total components of the partial typing morphisms in the original pairs.

\subsection{Compatibility with respect to typing}
\label{subsec:compatibility-typing}

In the same lines of typed graph morphisms where type compatibility must be ensured (see, e.g.,~\cite{ehrig2006fundamentals}) we have to make sure that graph chain morphisms, when needed, also respect typing.
For our MCMT rules, this means that $l: L \to I$ and $m: L \to S$ are compatible with respect to typing.
First we will discuss the typing compatibility of $l$, then we use an analogous explanation for $m$.

In case of $l: L \to I$, we do have actually a morphism $l: (L, \chainmorph[L]{}) \to (I, \chainmorph[I]{})$ where we require that, for all $0 < i \leq n$, the diagram (a) is commutative.
\begin{center}
	\begin{tikzpicture}[on grid,node distance=25mm]
	
	\tikzset{element/.style={inner sep=2pt,minimum height=1.3em,font=\itshape}}
	\tikzset{map/.style={->,>=stealth',semithick}}
	\tikzset{incmapl/.style={map, right hook-stealth'}}
	\tikzset{incmapr/.style={map, left hook-stealth'}}
	\tikzset{label/.style={auto,midway,font=\itshape,inner sep=2pt}}
	\tikzset{partmap/.style={map,postaction={decorate},decoration={markings,mark=at position 0.5 with {\draw circle [radius=.4ex];}}}}
	
	\node[element] (mm)							{$MM_i$};
	\node[element] (l)	[below left of=mm]		{L};
	\node[element] (i)	[below right of=mm]		{I};
	\node[element] (e1)	[below=13mm of mm]		{\(=\)};
	\node[element] (cap)	[below=9mm of mm]		{\((a)\)};
	
	\draw[partmap]	(l) to node[label]				(tl)	{\(\sigma^L_i\)}	(mm);
	\draw[partmap]	(i) to node[label,above right]	(ti)	{\(\sigma^I_i\)}	(mm);
	
	\draw[incmapl]	(l) to node[label,below]	(lid)	{\(l\)}	(i);
		
	\end{tikzpicture}
	\qquad\qquad
    \begin{tikzpicture}[on grid,node distance=25mm]
	
	\tikzset{element/.style={inner sep=2pt,minimum height=1.3em,font=\itshape}}
	\tikzset{map/.style={->,>=stealth',semithick}}
	\tikzset{incmapl/.style={map, right hook-stealth'}}
	\tikzset{incmapr/.style={map, left hook-stealth'}}
	\tikzset{label/.style={auto,midway,font=\itshape,inner sep=2pt}}
	
	\node[element] (mm)							{\chainname{MM}};
	\node[element] (l)	[below left of=mm]		{\(\chainname{L} = \domain{\sigma^L}\)};
	\node[element] (i)	[below right of=mm]		{\(\chainname{I} = \domain{\sigma^I}\)};
	\node[element] (l0)							{};
	\node[element] (e1)	[below=13mm of mm]		{\(=\)};
	\node[element] (cap)	[below=9mm of mm]		{\((b)\)};
	
	\draw[map]	(l) to node[label]				(tl)	{\((\sigma^L,id_{[n]})\)}	(mm);
	\draw[map]	(i) to node[label,above right]	(ti)	{\((\sigma^I,id_{[n]})\)}	(mm);
	
	\draw[incmapl]	(l) to node[label,below]	(lid)	{\((l,id_{[n]})\)}	(i);
	
	\end{tikzpicture}
\end{center}

As explained in~\cite{macias2017chains}, it is now straight forward to show that the family $l=(l_i:L_i \to I_i \mid i \in [n])$ of inclusion homomorphisms establishes a chain morphism $(l,id_{[n]}) : \cal{L}$ $\to \cal{I}$ (cf. Section~\ref{subsubsec:pushout-chain-equal-depth}).
The compatibility of typing for $l$ can be described by a commutative triangle (b) of chain morphisms.

Analogous to $l$, the type compatibility of $m: L \to S$ means that we require that, for all $0 < i \leq n$, the  diagram (c) below is commutative

\begin{center}
	\begin{tikzpicture}[on grid,node distance=25mm]
	
	\tikzset{element/.style={inner sep=2pt,minimum height=1.3em,font=\itshape}}
	\tikzset{map/.style={->,>=stealth',semithick}}
	\tikzset{label/.style={auto,midway,font=\itshape,inner sep=2pt}}
	\tikzset{partmap/.style={map,postaction={decorate},decoration={markings,mark=at position 0.5 with {\draw circle [radius=.4ex];}}}}
	
	\node[element] (mm)								{$MM_i$};
	\node[element] (tg)	[right of=mm]				{$TG_i$};
	\node[element] (l)	[below of=mm]				{L};
	\node[element] (s)	[below of=tg]				{S};
	\node[element] (eq)	[below right=18mm of mm]	{\(=\)};
	\node[element] (cap)	[below left=10mm of tg]	{\((c)\)};
		
	\draw[map]		(mm) to node[label]			(bf)	{\(b_i\)}				(tg);
	\draw[map]		(l) to node[label]			(mf)	{\(m\)}					(s);
	\draw[partmap]	(l) to node[label]			(tl)	{\(\sigma^L_i\)}		(mm);
	\draw[partmap]	(s) to node[label,right]	(ts)	{\(\sigma^L_{f(i)}\)}	(tg);
		
	\end{tikzpicture}
	\qquad\qquad
	\begin{tikzpicture}[on grid,node distance=25mm]
	
	\tikzset{element/.style={inner sep=2pt,minimum height=1.3em,font=\itshape}}
	\tikzset{map/.style={->,>=stealth',semithick}}
	\tikzset{label/.style={auto,midway,font=\itshape,inner sep=2pt}}
	
	\node[element] (mm)								{\chainname{MM}};
	\node[element] (tg)	[right of=mm]				{\chainname{TG}};
	\node[element] (l)	[below of=mm]				{\chainname{L}};
	\node[element] (s)	[below of=tg]				{\chainname{S}};
	\node[element] (eq)	[below right=18mm of mm]	{\(=\)};
	\node[element] (cap)	[below left=10mm of tg]	{\((d)\)};
	
	\draw[map]	(mm) to node[label]			(bf)	{\((b,f)\)}				(tg);
	\draw[map]	(l) to node[label]			(mf)	{\((m,f)\)}				(s);
	\draw[map]	(l) to node[label]			(tl)	{\((\sigma^L,id_{[n]})\)}	(mm);
	\draw[map]	(s) to node[label,right]	(ts)	{\((\sigma^S,id_{[m]})\)}	(tg);
	
	\end{tikzpicture}
	
\end{center}

Again, as explained in~\cite{macias2017chains}, it is now straight forward to show that the $m_i$ establish a chain morphism $(m, f) : \cal{L}$ $\to \cal{S}$, hence the compatibility of typing for $m$ can be described by a commutative square (d) of chain morphisms.

\subsection{Pushouts in the category $\cat{Chain}$}
\label{subsec:pushouts-category-chain}

As indicated in the previous section we use model transformations to define the behaviour of our MLM hierarchies. Since every branch in our MLM hierarchies is represented by a graph chain, the application of our model transformation rules will rely on pushout and pullback complement constructions in the category $\cat{Chain}$. Due to similarities between these constructions, the main focus of this section will be on the construction of the pushout of the following span
$$\xymatrix@R=7mm{
\chainname{S} && \chainname{L}\ar[ll]_{(m,f)} \ar@{^{(}->}[rr]^{(l,id_{[n]})} && \chainname{I} 
			}$$
of chain morphisms between inclusion chains in the category $\cat{Chain}$, where our assumption should ensure that the pushout becomes an inclusion chain as well. Moreover, the pushout in $\cat{Chain}$ should be fully determined by the construction of the pushout $D$ of the span
$$\xymatrix@R=7mm{
S_0=S && L_0=L\ar[ll]_{m_0} \ar@{^{(}->}[rr]^{l_0} && I_0=I
			}$$
in the category $\cat{Graph}$ of graphs and total graph homomorphisms. Especially, the multilevel typing of $D$, i.e., the level-wise pushouts of all the spans
$$\xymatrix@R=7mm{
S_{f(i)} && L_i\ar[ll]_{m_i} \ar@{^{(}->}[rr]^{l_i} && I_i \quad 1\leq i \leq n
			}$$
should be just parts of the pushout construction for the base $i=0$.

The result of the level-wise pushouts should be an inclusion chain of length $n+1$. 
The rule provides, however, only information about the typing at the levels $f([n])\subseteq[m]$. 
For the levels in $[m]\setminus f([n])$ we have to borrow the typing from the corresponding untouched levels in $\chainname{S}$. 
In terms of graph chain morphisms, this means that we factorize $(m,f)$ into two graph chain morphisms and that we construct the resulting inclusion chain in two pushout steps (see Figure~\ref{fig:two_po_construct_d}) where $\chainname{S}_{\downarrow f}=(\overline{S}_{\downarrow f},n,\tau^S_{\downarrow f})$ with
$\overline{S}_{\downarrow f}=[S_{f(n)}, S_{f(n-1)},\ldots,S_{f(1)},S_0]$ and
\(\tau^S_{\downarrow f} = (\tau^S_{f(j),f(i)}:\graphname{S}{f(j)} \partialmap \graphname{S}{f(i)}\mid 0\le\level{i}<\level{j}\le\level{n})\). 

\begin{figure}[h!]
\begin{center}
	\begin{tikzpicture}[on grid,node distance=25mm]
	
	\tikzset{element/.style={inner sep=2pt,minimum height=1.3em,font=\itshape}}
	\tikzset{map/.style={->,>=stealth',semithick}}
	\tikzset{incmapl/.style={map, right hook-stealth'}}
	\tikzset{incmapr/.style={map, left hook-stealth'}}
	\tikzset{label/.style={auto,midway,font=\itshape,inner sep=2pt}}
	
	\node[element] (l)										{\chainname{L}};
	\node[element] (sf)	[right=40mm of l]					{\(\chainname{S}_{\downarrow f}\)};
	\node[element] (s)	[right=40mm of sf]					{\chainname{S}};
	\node[element] (i)	[below of=l]						{\chainname{I}};
	\node[element] (ug)	[below of=sf]						{\(\chainname{D}_{\downarrow f}\)};
	\node[element] (si)	[below of=s]						{\(\chainname{D}\)};
	\node[element] (eq)	[below right=12mm and 20mm of l]	{\((1)\)};
	\node[element] (eq)	[below right=12mm and 25mm of sf]	{\((2)\)};
	\draw[map]		(l) to node[label]			(mu1)	{\((m,id_{[n]})\)}								(sf);
	\draw[map]		(sf) to node[label]			(id1)	{\((\overline{id},f)\)}								(s);
	\draw[incmapl]	(l) to node[label,left]		(in1)	{\((l,id_{[n]})\)}								(i);
	\draw[incmapl]	(sf) to node[label]			(in2)	{\((s_{\downarrow f},id_{[n]})\)}								(ug);
	\draw[incmapl]	(s) to node[label]			(in3)	{\((s,id_{[m]})\)}								(si);
	\draw[map]		(i) to node[label,below]	(mu2)	{\((d,id_{[n]})\)}	(ug);
	\draw[incmapl]	(ug) to node[label,below]	(id2)	{\((\overline{id},f)\)}								(si);
		
	\end{tikzpicture}
\end{center}
\caption{Two pushout steps to construct the inclusion chain $\chainname{D}$}
\label{fig:two_po_construct_d}
\end{figure}

The graph chain morphism $(\overline{id},f):\chainname{S}_{\downarrow f}\to\chainname{S}$ is a level-wise identity and just embeds a chain of length $n$ into a chain of length $m$, i.e., $\overline{id} = (id_{f(i)}: S_{f(i)} \to S_{f(i)} \mid i \in [n] )$. $\chainname{D}_{\downarrow f}$ and $(\overline{id},f):\chainname{D}_{\downarrow f}\to\chainname{D}$ are defined analogously.

In the first step (1) we construct the pushout of inclusion chains of equal depth and in pushout step (2) we just exchange some levels in $\chainname{S}$ by the corresponding extended levels provided by $\chainname{D}_{\downarrow f}$.

\subsubsection{Pushouts for chains with equal depth}
\label{subsubsec:pushout-chain-equal-depth}

We describe now the construction of the pushout (1) of chains of graphs of equal length (see Figure~\ref{fig:two_po_construct_d}).
The rest of this section depends on the special pushout construction in category \cat{Graph} for a span with one inclusion graph homomorphism and a necessary result about the structure of special mediating morphisms.
A recapitulation of this construction is presented in~\cite{macias2017chains}.

We construct the corresponding pushout of graph homomorphisms (in category \cat{Graph}) for each level $i \in [n]$ as follows.

\begin{center}
	\begin{tikzpicture}[on grid,node distance=25mm]
	
	\tikzset{element/.style={inner sep=2pt,minimum height=1.3em,font=\itshape}}
	\tikzset{map/.style={->,>=stealth',semithick}}
	\tikzset{label/.style={auto,midway,font=\itshape,inner sep=2pt}}
	\tikzset{incmapl/.style={map, right hook-stealth'}}
	\tikzset{incmapr/.style={map, left hook-stealth'}}
	
	\node[element] (gi)								{\(L_i\)};
	\node[element] (hi)	[right=40mm of gi]				{\(I_i\)};
	\node[element] (ki)	[below of=gi]				{\(S_{f(i)}\)};
	\node[element] (pi)	[below of=hi]				{\(D_{f(i)} := S_{f(i)} + I_i \setminus L_i\)};
	\node[element] (po) [below right=13mm and 18mm of gi]	{PO};
	
	\draw[incmapl] (gi) to	node[label]			(ph)	{\(l_i\)}	(hi);
	\draw[incmapl] (ki) to	node[label]			(phs)	{\(s_{f(i)}\)}	(pi);
	\draw[map] (gi) to	node[label,left]	(ps)	{\(m_i\)}	(ki);
	\draw[map] (hi) to	node[label]			(pss)	{\(d_i := m_i + id_{I_i\setminus L_i}\)}	(pi);
	
	\end{tikzpicture}
\end{center}

We consider any level $1\leq i \leq n$ together with the base level $0$. Since $I_i \sqsubseteq I_0$, we have that $I_i \setminus L_i \sqsubseteq I_0 \setminus L_0$ and thus we get a span of pullbacks.

\begin{center}
	\begin{tikzpicture}[on grid,node distance=25mm]
	
	\tikzset{element/.style={inner sep=2pt,minimum height=1.3em,font=\itshape}}
	\tikzset{map/.style={->,>=stealth',semithick}}
	\tikzset{incmapl/.style={map, right hook-stealth'}}
	\tikzset{incmapr/.style={map, left hook-stealth'}}
	\tikzset{label/.style={auto,midway,font=\itshape,inner sep=2pt}}
	
	\node[element] (l)										{$S=S_0$};
	\node[element] (sf)	[right=40mm of l]					{\(D:=D_0\)};
	\node[element] (s)	[right=40mm of sf]					{$I=I_0$};
	\node[element] (i)	[below of=l]						{$S_{f(i)}$};
	\node[element] (ug)	[below of=sf]						{\(D_{f(i)}\)};
	\node[element] (si)	[below of=s]						{\(I_i\)};
	
	\draw[incmapl]		(l) to node[label]			(mu1)	{\(s:=s_0\)}								(sf);
	\draw[map]		(s) to node[label,above]			(id1)	{\(d:=d_0\)}								(sf);
	\draw[incmapl]	(i) to node[label,left]		(in1)	{\(\tau^S_{i,0}\)}								(l);
	\draw[map]	(ug) to node[label]			(in2)	{\(\tau^D_{i,0} := \tau^S_{i,0} + \tau^I_{i,0} \setminus \tau^L_{i,0} \)}								(sf);
	\draw[incmapl]	(si) to node[label,midway,right]			(in3)	{\(\tau^I_{i,0} := \tau^L_{i,0} + \tau^I_{i,0} \setminus \tau^L_{i,0}\)}								(s);
	\draw[incmapl]		(i) to node[label,below]	(mu2)	{\(s_{f(i)}\)}	(ug);
	\draw[map]	(si) to node[label,below]	(id2)	{\(d_i\)}								(ug);
		
	\end{tikzpicture}
\end{center}

The sequence $[D_{f(n)}, D_{f(n-1)}, \dots, D_{f(1)}, D_0]$ defines, according to Corollary~\ref{cor:refactoring}, an inclusion chain that we denote by $\chainname{D}_{\downarrow f}$.
To show that the family of inclusion graph homomorphisms $(s_{f(i)}: S_{f(i)} \hookrightarrow D_{f(i)}  \mid i \in [n])$ defines a graph chain morphism $(s_{\downarrow f}, id_{[n]}) : \chainname{S}_{\downarrow f} \hookrightarrow \chainname{D}_{\downarrow f}$ we have to show, according to Definition~\ref{def:graph-chain-morphism}, that we have for any $0 \leq i < j \leq n$ a commutative diagram with a pullback as follows. 

\begin{center}
			$\xymatrix@R=7mm{
				S_{f(i)}  \ar[rr]^{f_{f(i)}}
				&& D_{f(i)}\\
				S_{f(j)} \cap S_{f(i)} \ar@{_{(}->}[d]_-{\sqsubseteq} \ar@{^{(}->}[rr]^-{\sqsubseteq} \ar@{^{(}->}[u]^-{\sqsubseteq} \ar@{}[drr]|{PB}
				&& D_{f(j)} \cap D_{f(i)} \ar@{_{(}->}[d]_-{\sqsubseteq} \ar@{^{(}->}[u]^{\sqsubseteq} \\
				S_{f(j)} \ar[rr]^{\sigma^S_j}
				&& D_{f(j)} \\
			}$\\
	\end{center}

Analogously, it can be shown that the family $(d_{i}: I_{i} \hookrightarrow D_{f(i)}  \mid i \in [n])$ establishes a chain morphism $(d, id_{[n]}) : \chainname{I} \hookrightarrow \chainname{D}_{\downarrow f}$.
Since the resulting commutative square (1) of chain morphisms (see Figure~\ref{fig:two_po_construct_d}) is obtained by level-wise pushout constructions, it is straight forward to show that (1) becomes indeed a pushout in \cat{Chain}.

\subsubsection{Pushout by extension}
\label{subsubsec:pushout-by-extension}

In this subsection we will define the pushout construction (2) in Figure~\ref{fig:two_po_construct_d}.
To obtain an inclusion chain $\chainname{D}$ of length $m+1$ we fill the gaps in the sequence $[D_{f(n)}, D_{f(n-1)}, \dots, D_{f(1)}, D_0]$ of subgraphs of $D=D_0$, constructed in Section~\ref{subsubsec:pushout-chain-equal-depth}, by corresponding subgraphs of $S$ from the sequence $[S_m, S_{m-1}, \dots, S_1, S_0]$.
For any $j \in [m] \setminus f([n])$ we denote by $f^{-}(j)$ the unique index in $[n]$ such that $f(f^-(j))=j$.
We define the sequence $\bar{D}$ of subgraphs of $D=D_0$ as follows.
\begin{equation*}
	D_j := \begin{cases}
		D_j &\text{if $j \in f([n])$}\\ 
		S_j &\text{if $j \in [m] \setminus f([n])$}
	\end{cases}
\end{equation*}
and denote by $\chainname{D} = \chain{D}{m}{\tau^D}$ the corresponding inclusion chain according to Corollary~\ref{cor:refactoring}.
The family $\bar{id} = (id_{D_{f(i)}}: D_{f(i)} \to D_{f(i)} \mid i \in [n])$ of graph homomorphisms defines trivially a chain morphism $(\bar{id},f) : \chainname{D}_{\downarrow f} \to \chainname{D}$.

It remains to show that the family $s = (s_j: S_j \to D_j \mid j \in [m])$ of graph homomorphisms defined by 
\begin{equation*}
	s_j := \begin{cases}
		s_j : S_j \hookrightarrow D_j &\text{if $j \in f([n])$}\\ 
		id_{S_j}:  S_j \to D_j = S_j &\text{if $j \in [m] \setminus f([n])$}
	\end{cases}
\end{equation*}
establishes a chain morphism $(s, id_{[m]}) : \chainname{S} \to \chainname{D}$.

\begin{wrapfigure}[14]{r}[0mm]{.55\textwidth}
	\centering
\begin{center}
	\begin{tikzpicture}[on grid,node distance=30mm]	
	\tikzset{element/.style={inner sep=2pt,minimum height=1.3em,font=\itshape}}
	\tikzset{map/.style={->,>=stealth',semithick}}
	\tikzset{label/.style={auto,midway,font=\itshape,inner sep=2pt}}
	
	\node[element] (l)										{\chainname{L}};
	\node[element] (s)	[right of=l]						{\chainname{S}};
	\node[element] (i)	[below of=l]						{\chainname{I}};
	\node[element] (d)	[right of=i]						{\chainname{D}};
	\node[element] (mm)	[below right=10mm and 25mm of i]	{\chainname{MM}};
	\node[element] (tg)	[right of=mm]						{\chainname{TG}};
	
	\draw[map]			(l) to node[label]				(mf)	{\((m,f)\)}					(s);
	\draw[map] 			(l) to node[label,left]			(li)	{\((l,id_{[n]})\)}			(i);
	\draw[map] 			(s) to node[label,left]			(sd)	{\((s,id_{[m]})\)}			(d);
	\draw[map,dotted]	(d) to node[label,left]			(dt)	{\((\sigma^D,id_{[m]})\)}	(tg);
	\draw[map]			(i) to node[label,below left]	(im)	{\((\sigma^I,id_{[n]})\)}	(mm);
	\draw[map]			(s) to node[label,above right]	(st)	{\((\sigma^S,id_{[m]})\)}	(tg);
	\draw[map]			(l) to node[label,above right, near start]	(lm)	{\((\sigma^L,id_{[n]})\)}	(mm);
	\draw[map]			(i) to node[label,below]		(df)	{\((d,f)\)}					(d);
	\draw[map]			(mm) to	node[label,below]		(bf)	{\((b,f)\)}					(tg);	
	\end{tikzpicture}
\end{center}
\end{wrapfigure}

The resulting square (2) in Figure~\ref{fig:two_po_construct_d} is commutative by construction and it is straightforward to show that it becomes a pushout in \cat{Chain} as well.
Composing the two pushouts (1) and (2) we obtain a pushout in \cat{Chain} that provides us, finally, with a chain morphism from \chainname{D} to \chainname{TG} that materialises the required multilevel typing of the constructed graph $D$.
The left triangle commutes due to compatible typing of the rule.
The roof square commutes since the match \((m,f)\) is type compatible.
This gives us \((m,f);(\sigma^S,id_{[m]}) = (l,id_{[n]});(\sigma^I,id_{[n]});(b,f)\), thus the pushout universal property of the back of the square gives us a unique chain morphism \((\sigma^D,id_{[m]}): \chainname{D} \to \chainname{TG}\) such that the two desired typing compatibilities \((s,id_{[m]});(\sigma^D,id_{[m]}) = (\sigma^S,id_{[m]})\) and \((d,f);(\sigma^D,id_{[m]}) = (\sigma^I,id_{[n]});(b,f)\) are satisfied.

\section{Tooling}
\label{sec:tooling}

We present in this section our current results towards the full support for multilevel modelling and MCMTs. 
Specifically, we introduce in Section~\ref{subsec:textual-dsl} our textual language for the specification of MCMTs --- equivalent to the graphical representation used in previous sections --- and its editor. 
We present a proliferation algorithm that generates two-level MTs out of given set of MCMTs, which can then be used on conventional model transformation engines.
This way, we exploit the advantages of MCMTs over two-level MTs --- as presented in Section~\ref{subsec:why-using-mcmt} --- but avoid the necessity of creating a brand new transformation engine for MCMTs.
A full-fledged implementation of a MLM transformation engine is left as future work.
These tools are available as Eclipse plugins from the MultEcore website\footnote{MultEcore website: \url{http://ict.hvl.no/multecore}.}.

\subsection{Textual DSL for MCMTs}
\label{subsec:textual-dsl}

\begin{figure}
	\centering
	\includegraphics[width=\textwidth]{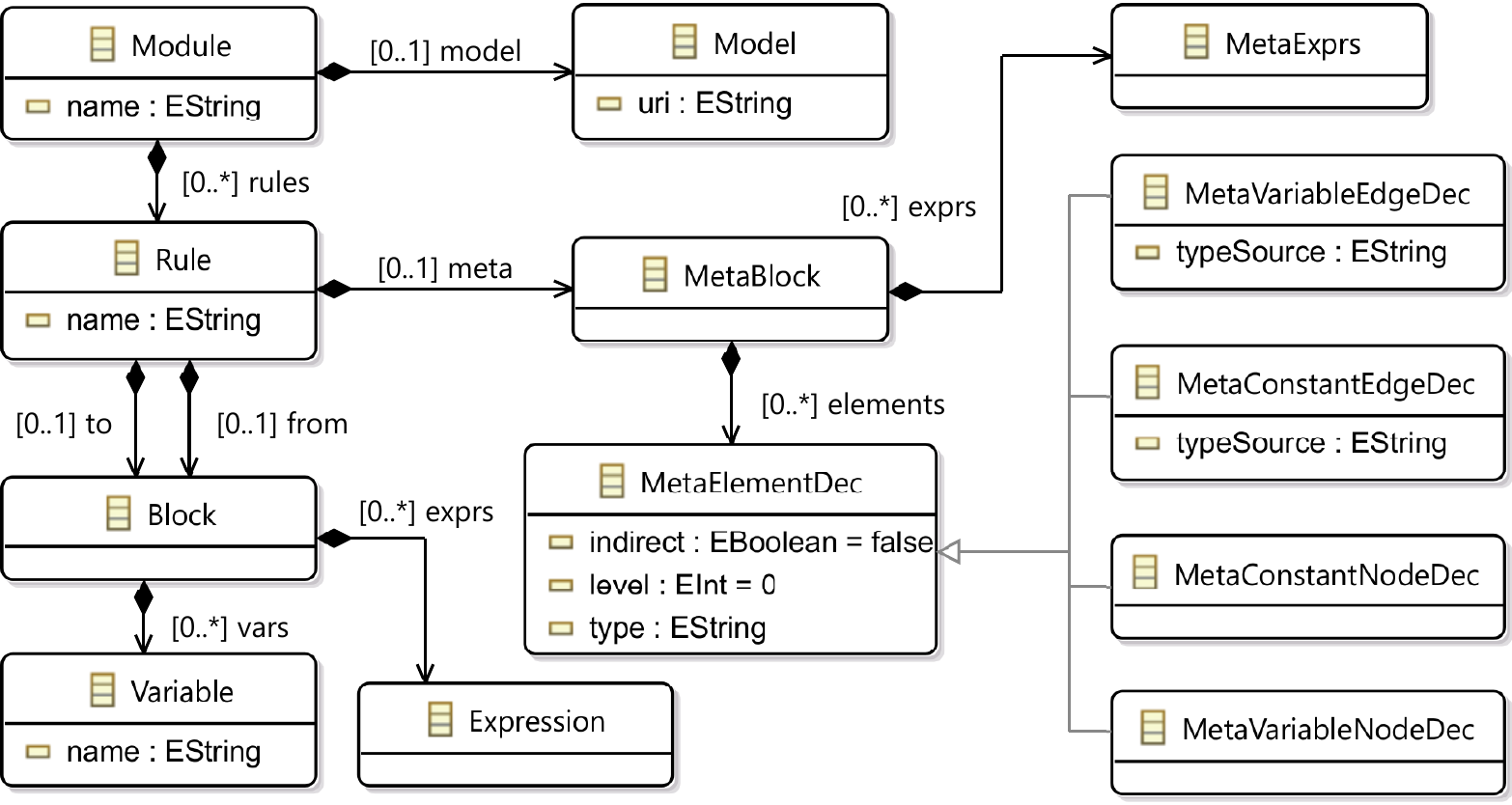}
	\caption{Fragment of the metamodel defining the abstract syntax of our textual DSL}
	\label{fig:textual-dsl-metamodel}
\end{figure}

The abstract syntax of our DSL for MCMTs is defined as an Ecore metamodel, partially depicted in Figure~\ref{fig:textual-dsl-metamodel}.
This metamodel has been used to create an editor using Xtext~\cite{xtext} which provides syntax highlighting, error detection, code completion and outline visualization.

This DSL provides the specification of modules containing a collection of multilevel coupled transformation rules defined independently from a hierarchy, but that will be matched against one during the proliferation process (see Section~\ref{subsec:proliferation}).
Therefore, the MCMTs are applied to a single hierarchy in each execution.

\begin{figure}
	\centering
	\includegraphics[width=.7\textwidth]{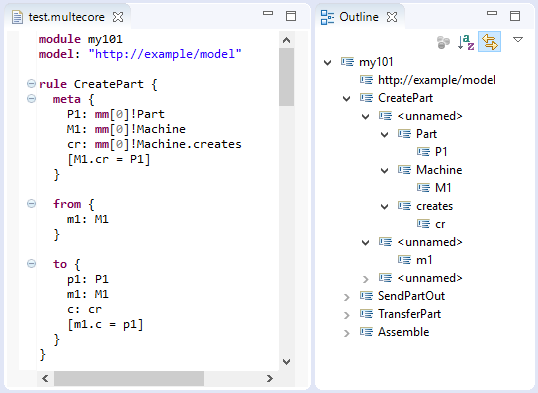}
	\caption{Screenshot of the editor with the textual representation of the \emph{CreatePart} rule}
	\label{fig:textual-editor-screenshot}
\end{figure}

In order to briefly present this DSL, we use the textual representation of the MCMT rule \textit{CreatePart}, already introduced in Section~\ref{subsec:why-using-mcmt}.
This textual version is shown in Figure~\ref{fig:textual-editor-screenshot}, from a screenshot of our Eclipse-based editor.
As stated in Section~\ref{subsec:why-using-mcmt}, \textsf{Rule} elements are structured into three organizational components, namely META, FROM and TO, named in lowercase in the textual syntax.
These blocks contain graph pattern declarations, plus expressions that relate the elements to each other.
The \textsf{meta} block must contain a valid pattern, but the \textsf{from} and \textsf{to} blocks may be empty.
In the \textsf{meta} block, a pattern is specified by means of \textsf{MetaElementDec}, used to declare both nodes and edges, which can in turn be constant or not.
Also, this block may contain assignment expressions, used in this example to specify the structural relationships between the declared nodes by means of the declared edges.
In the example, \textsf{P1}, \textsf{M1} are declared as node variables, while \textsf{creates} is declared as a variable edge.
Specifying any of these as a constant would just require to add the character \textsf{\$} as a suffix to their type.
Note that the type names of the \textsf{meta} block elements are suffixed by the keyword \textsf{mm} followed by the level (index) at which these types is actually declared, starting from \level{0} and increasing downwards.
Note that this editor assumes Ecore at level \level{0}, which does not have to be explicitly given.

In the \textsf{from} and \textsf{to} blocks we can define patterns according to the \textsf{MetaElementDec} previously declared in the \textsf{meta} part.
In the example, the \textsf{from} block of the rule defines a pattern consisting of just one variable \textsf{m1}, while its \textsf{to} block comprises three variable declarations and an assignment expression between them.
This pattern mirrors precisely the one depicted in Figure~\ref{fig:pls-rule-create-part}.

\subsection{Proliferation process}
\label{subsec:proliferation}

As mentioned before, an MCMT can generically, but precisely, comprise a number of two-level specific rules which are applicable to just one given source model.
For matching an MCMT in a modelling hierarchy and executing it, we can consider two different approaches.

A first approximation requires the implementation of a custom execution engine that tackles MCMTs and multilevel modelling stacks.
This kind of approach would offer a more seamless execution of MCMTs in an integrated enviroment, also facilitating possible extensions and eliminating the necessity of back and forth translation of hierarchies and MCMTs to existing transformation engines.
However, the cost of implementing such a custom engine from scratch, including the re-implementation of basic functionalities which already exist in transformation engines, would be too high.
Moreover, it might lock the transformation engine to a single MLM-tool.

As an alternative, it is possible to perform a pre-processing step that can automatically generate the operationally equivalent two-level rules which, while correct, are in general too numerous and cumbersome to specify by hand.
This approach can be combined with other well-established, well-understood and optimised transformation engines like Henshin~\cite{ArendtBJKT10}, Groove~\cite{Rensink03,ghamarian2012groove}, ETL~\cite{KolovosPP08} or ATL~\cite{JouaultABK08}.
In other words, this process of automatic proliferation results in a more flexible implementation, in which we do not even strictly depend on a particular transformation engine.
Besides, the process of adapting both the target hierarchy and the MCMTs to a two-level setting can be done automatically, and indeed needs to be performed only once before running the whole scenario.

Specifically, our automatic proliferation procedure consists in the generation of two-level model transformations for every possible match of the variables defined in an MCMT.
The goal of this process is to replace the variability in the types of the elements in the FROM and TO blocks --- which cannot be interpreted by mainstream model transformation engines --- by specific types, so that we can generate automatically an executable set of rules for any two-level engine.

Once specified, an MCMT is run by specifying in which model it will be applied, as in any model transformation engine.
The specification of the input model, which is part of a multilevel hierarchy, uniquely identifies its branch in the hierarchy tree.
That is, we can ignore at execution time the same-level sibling models of the input one as well as the sibling models of any metamodel of the input model in the levels above.
Furthermore, in the exceptional case that the input model is not in the bottommost level of the hierarchy, all models in the levels below it can be ignored for the execution.
As a consequence of these simplifications, the relevant part of the hierarchy will always be a stack of models, without any branches.
Hence, we will hereafter use the term \emph{stack} instead of \emph{hierarchy}.

The way to replace the variable types of the MCMT by specific ones, is by looking at the META part of the MCMT rule and finding all possible matches of its variables into the types defined in the modelling stack.
Looking back to Figure~\ref{fig:multilevel-coupled-rule-formal}, this process will establish all the possible maps \textit{b\textsubscript{x}}, and generate one two-level rule for each valid combination that involves one map per level in the META part, which is represented as the sequence of graphs \textit{MM\textsubscript{x}}.
Once proliferated, it is necessary to perform a second matching process \textit{m} between the FROM graph \textit{L} and the input model \textit{S}, in order to actually apply the --- now proliferated --- rule.
Such operation will be performed by the selected two-level engine, so both \textit{L} and \textit{S} are excluded from the proliferation algorithm.

The core of our MCMT proliferation mechanism is a matching algorithm between the meta part of an MCMT rule and the stack of metamodels of the input model. 
We describe our matching algorithm in the following sections.

\subsection{Matching algorithm}
\label{subsec:matching-algorithm}

The algorithm that generates the list of all possible values of the variables that yield valid two-level MTs is shown in Algorithm~\ref{alg:matching}.
It consists of a recursive function with four input parameters and an input/output parameter.

\begin{algorithm}[tb]
	\footnotesize
    \caption{Matching algorithm}
    \label{alg:matching}
    \begin{algorithmic}[1] 
      \Procedure{Match}{$MM,TG,\mathit{mmLevel},\mathit{tgLevel},matches$}
        \If{$\mathit{mmLevel} = MM.size$}
          \State \textbf{return} $true$ \Comment{End of pattern reached}
        \EndIf
        \State $\mathit{found} \gets false$
        \While{$\mathit{tgLevel}<TG.size$} \Comment{Every level in hierarchy}
          \State $\mathit{maps \gets graphMatch(MM[mmLevel],TG[tgLevel],matches[matches.size - 1])}$ \label{alg:matching:graphmatch}
          \ForAll {$m \in maps$}
            \State $size \gets matches.size$
            \If{$size > 0\ \textbf{and}\ \mathit{mmLevel} > 0$}
              \State $\mathit{currentMatch} \gets matches[size - 1]$ \label{alg:matching:current-1}
              \State $matches[size - 1] \gets \mathit{currentMatch} \cup m$ \label{alg:matching:union-m}
              \State $\mathit{found} \gets \mathit{found}\ \textbf{or}\ match(MM,TG,\mathit{mmLevel}+1,\mathit{tgLevel}+1,matches)$
              \State $matches[matches.size] \gets \mathit{currentMatch}$ \label{alg:matching:current-2}
            \Else
              \State $matches[size] \gets m$
              \State $\mathit{found} \gets \mathit{found}\ \textbf{or}\ match(MM,TG,\mathit{mmLevel}+1,\mathit{tgLevel}+1,matches)$
            \EndIf
          \EndFor
          \State $\mathit{tgLevel} \gets \mathit{tgLevel} + 1$
        \EndWhile
        \If{$\mathit{mmLevel} > 0$}
          \State $matches[matches.size-1] \gets \emptyset$  \Comment{Remove incomplete match}
        \EndIf
        \State \textbf{return} $\mathit{found}$ \label{alg:matching:return}
      \EndProcedure
    \end{algorithmic}
\end{algorithm}

The first two input parameters, \algstyle{MM} and \algstyle{TG}, represent all the metalevels of an MCMT depicted in Figure~\ref{fig:multilevel-coupled-rule-formal} as \textit{MM\textsubscript{x}} and \textit{TG\textsubscript{y}}.
Hence, they represent the META part of an MCMT (the typing chain \algstyle{MM}) and the levels above the one were the MCMT will be applied (\algstyle{TG}).
That is, \algstyle{MM} represents the \textit{pattern} that the algorithm tries to match and \algstyle{TG} represent the \textit{stack} against which the algorithm tries to match \algstyle{MM}.
Both inputs are accessed level by level, in all valid combinations, as the algorithm progresses.
This way, it establishes all possible \textit{maps} between pairs of levels that conform a full \textit{match} --- and hence result in a proliferated rule.

The next two parameters, \algstyle{mmLevel} and \algstyle{tgLevel}, are the indexes used to indicate the current levels that are being accessed in \algstyle{MM} and \algstyle{TG}, respectively.
In the first call, both have value zero, which will increase as the algorithm progresses.
Since we number the levels of both the \algstyle{MM} and \algstyle{TG} increasing downwards, the matches will be generated in a top-down manner.
This decision is not arbitrary: the match between two levels depends, among other things, of the types of their elements; since those types may be other elements in levels above, these need to be matched first, so that we can check that the types of any two elements are compatible --- that is, the type in the pattern has been previously mapped to the type in the stack.

The last parameter, \algstyle{matches}, contains the list of all valid matches found by the algorithm.
It is initially empty, and it is required to work as an input/output variable (by reference) so that the algorithm can modify it in each recursive call and, afterwards, propagate the result of such calls to provide the final list.
For the same reason described in the previous paragraph, the information of the maps established so far for any match needs to be passed to \algstyle{graphMatch} so that it can be taken into account to produce only those maps that are consistent with the maps of the types.

The algorithm depends on an auxiliary function \algstyle{graphMatch} (line~\ref{alg:matching:graphmatch}) that calculates the possible matches between two models.
This auxiliary algorithm is described in Section~\ref{subsec:graph-matching}.

The function returns a boolean value that indicates whether the algorithm has found valid matches or not.
In the latter case, the rule would not be applicable and, consequently, would not be proliferated.
This value is also used by the algorithm to filter all potential matches that are discarded, in case they could not be completed.

The algorithm works as follows.
First, the base case of the recursion is triggered if the \algstyle{mmLevel} has reached the size of the pattern of the rule (\algstyle{MM}).
This would indicate that the algorithm has found a sequence of partial matches all the way down to the last level of \algstyle{MM}, which constitute a full match.
If that is not the case, the first time that a pair of levels is explored as part of one sequence of matches, the flag \algstyle{found} is set to false.

The core of the algorithm is a while loop that attempts to match the current \algstyle{mmLevel} with all the remaining \algstyle{tgLevel}s.
The first value is the one which was passed to the function in the call, which will increase by $1$ in each iteration.
On each iteration, the list of maps that are possible to establish for \algstyle{mmLevel} and \algstyle{tgLevel} are calculated with the call to \algstyle{graphMatch}.
This function takes the two levels of \algstyle{MM} and \algstyle{TG} indicated by the indexes, and gives back a list of maps from all the elements in the pattern into elements of the multilevel stack.
This results are stored in the \algstyle{maps} variable, and then iterated in the for loop that performs the recursive calls in order to complete the match.

Inside the for loop, the current match \algstyle{m} is stored in \algstyle{matches}, and the recursive call is made afterwards.
However, a case distinction is required.
In those cases where both the size of the \algstyle{matches} variable and \algstyle{mmLevel} are not zero, these values indicate that the algorithm has already found some maps for the current match.
In that case, such information is stored in the last position of the list (\algstyle{matches[size-1]}), and \algstyle{m} is simply added to that position (line~\ref{alg:matching:union-m}).
Furthermore, in case more than one match can be generated from the current sequence of maps and other values of \algstyle{m}, that current information is temporarily stored in \algstyle{currentMatch} and duplicated in the last position of \algstyle{matches} after the recursive call (lines~\ref{alg:matching:current-1} and~\ref{alg:matching:current-2}).
In the cases where the \algstyle{size} of \algstyle{matches} is zero (empty list) or \algstyle{mmLevel} is zero, there is no partial information to be preserved before storing \algstyle{m} and making the recursive call, so \algstyle{m} is simply added to the tail of the list, since it is the first map of the sequence.
In both cases, the recursive call is performed in the same way, and its return value is used to indicate if the match was completed after the current map or not.
The result, stored in \algstyle{found}, will be then propagated to the calling function in the same way (line~\ref{alg:matching:return}).

After all iterations of a loop are completed, and due to the duplication described above when \algstyle{mmLevel} and \algstyle{size} are not zero, the last partial result that was not completed is taken away from the list of matches.

Note that \algstyle{mmLevel} and \algstyle{tgLevel} are increased in such a way that it always holds that \(\algstyle{mmLevel}\leq\algstyle{tgLevel}\).
Moreover, every time that \algstyle{mmLevel} is increased, the same happens to \algstyle{tgLevel}.
This ensures that the sequence of graph matches are consistent with their depiction in Figure~\ref{fig:multilevel-coupled-rule-formal}, where two \textit{b\textsubscript{x}} cannot have the same source or target, or ``cross'' each other.

To sum up, the algorithm is first called with the \algstyle{MM} specified in the META part of the MCMT, the \algstyle{TG} that represent the multilevel stack on top of the model to be transformed, both \algstyle{mmLevel} and \algstyle{tgLevel} initialized at zero and \algstyle{matches} initialized empty.
It will recursively explore all valid combination of maps that cover all the levels in the pattern, and return those in \algstyle{matches}, plus a boolean value set to false if no full matches were found.

\begin{figure}[ht]
	\centering
	\includegraphics[width=.7\linewidth]{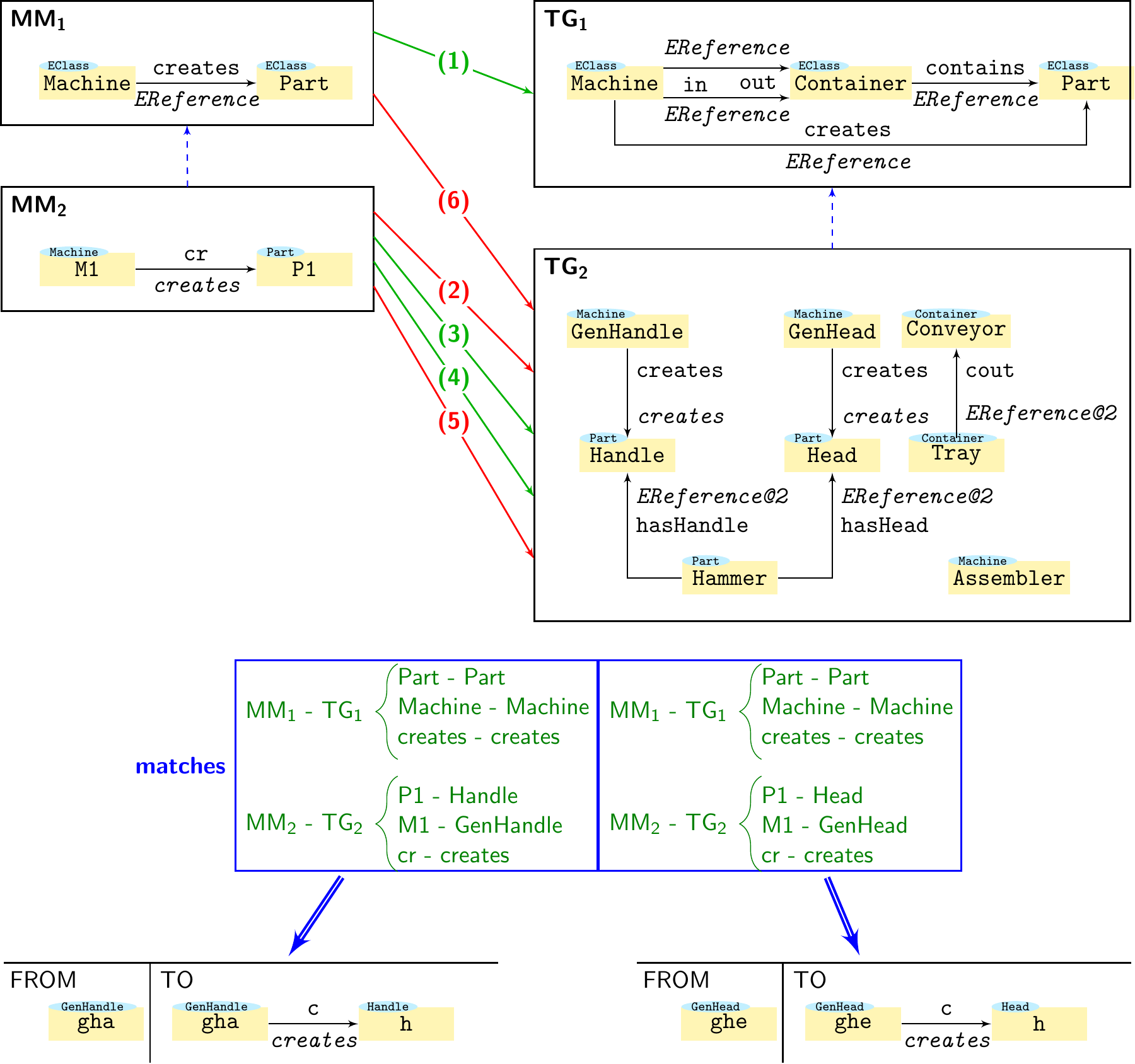}
	\caption{Example of application of the matching algorithm and subsequent proliferation applied to the rule \emph{CreatePart}}
	\label{fig:proliferation-example-overview}
\end{figure}

Figure~\ref{fig:proliferation-example-overview} displays an example of the application of Algorithm~\ref{alg:matching} to the MCMT \emph{CreatePart} that we previously presented in Figure~\ref{fig:pls-rule-create-part}.
Following the top-down mechanism we already described, the two \textit{MM\textsubscript{x}} representing the META part of the rule are matched against the multilevel stack in top of the model where the rule is applied.
The rule \emph{CreatePart} is applied in this example to the model \elementname{hammer\_config} in Figure~\ref{fig:pls-multilevel-hierarchy}d.
Hence, the stack of models \textit{TG\textsubscript{y}} consists of \elementname{generic\_plant} and \elementname{hammer\_plant} --- Figure~\ref{fig:pls-multilevel-hierarchy}(a) and (b), respectively.
For simplicity reasons, the potencies of all elements, as well as the multiplicities of relations, are excluded from the diagram, since they are not specified and hence make no difference in the execution of the algorithm.

\subsection{Graph matching}
\label{subsec:graph-matching}

The match between a graph defined in the META part of an MCMT and a subgraph of one of the models in the multilevel stack is done by means of graph homomorphisms, plus some restrictions.
The algorithm for graph matching is a modification of the Ullman algorithm~\cite{Ullmann76}, as proposed in~\cite{saltz2013fast}.
Basically, we take into account some modelling aspects in order to adapt the process from pure graphs to modelling.

Recall that nodes within a graph are uniquely identified by their name, which act as unique identifiers.
As for arrows, names can be reused as long as the source or target nodes --- at least one --- are different.
Moreover, both nodes and arrows are fundamentally defined by their types.
Hence, in order to match any element, it is required that the one in the pattern and its counterpart in the hierarchy have matching types.
This restriction does not conflict with the possibility of having variable types ---
see, e.g., rules \textit{Assemble} (Figure~\ref{fig:pls-rule-assemble}) and \textit{TransferPart} (Figure~\ref{fig:pls-rule-transfer-part}).

As shown in Section~\ref{subsec:proliferation}, the proliferation algorithm works in a top-down manner, so that before matching any element, its type has already been matched, and it is possible to use that information for the current match.

For arrows, where multiplicity is part of their definition, the lower and upper bounds can also be used for matching purposes, taking into account that a more restrictive multiplicity in the pattern will match a less restrictive one in the hierarchy.
For example, a pattern arrow with multiplicity \elementname{1..2} will match a hierarchy arrow with multiplicity \elementname{0..3} if the rest of the matching conditions are met.

In addition, the notion of potency can also be used in the match if required, although it is ignored if not specified.
In a similar manner to arrow multiplicity, the less restrictive ones in the hierarchy --- broader intervals --- can be matched by more restrictive patterns --- narrower intervals.

Lastly, the distinction between variables and constants, already presented in Section~\ref{subsec:mcmt-pls}, also influences the graph matching algorithm.
In this case, the name of the element is taken into account for the matching.
For nodes, the algorithm must find a corresponding node with all the restrictions aforementioned, plus an equal name, in order to get a successful match.
For arrows, the type is still required, since the name does not necessarily identifies a single arrow.
To properly identify one specific arrow, we need to identify its source and target nodes as constants too.

\begin{figure}[ht]
	\centering
	\includegraphics[width=.7\linewidth]{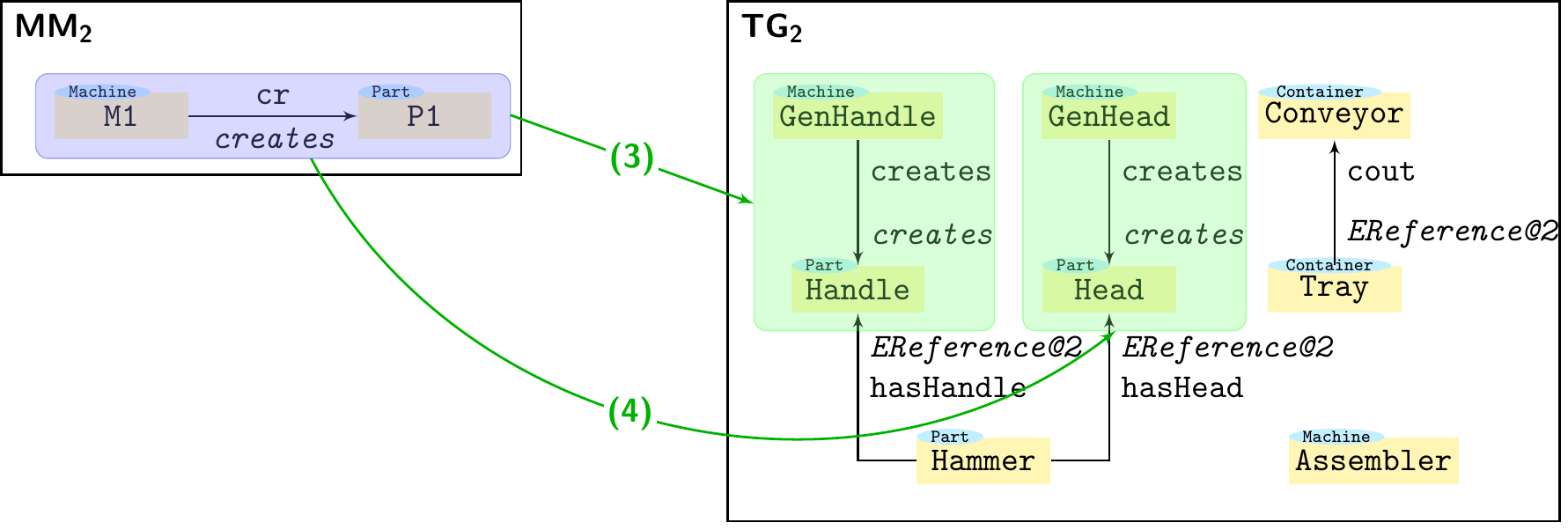}
	\caption{Detail of the graph matching process on the rule \emph{CreatePart}}
	\label{fig:proliferation-example-detail}
\end{figure}

Figure~\ref{fig:proliferation-example-detail} displays a fragment of Figure~\ref{fig:proliferation-example-overview}, in which, after matching \elementname{MM\textsubscript{1}} to \elementname{TG\textsubscript{1}}, the proliferation algorithm proceeds to find matches between \elementname{MM\textsubscript{2}} to \elementname{TG\textsubscript{2}}.
As depicted, the pattern (blue) can match two different subgraphs (green), resulting in two different proliferated rules.
This can be achieved since the proliferation process has already matched, before a recursive call, the elements \elementname{Machine}, \elementname{Part} and \elementname{creates} in \elementname{MM\textsubscript{1}} to their homonyms in \elementname{TG\textsubscript{1}}.

\section{Related Work}
\label{sec:related-work}

We discuss in this section works related to our proposal in the fields of multilevel modelling and on the use of model transformations to define the behaviour of languages and systems.

\subsection{Multi-level metamodelling}

Multi-level metamodelling was initially proposed by Atkinson and K\"uhne~\cite{atkinson1997distributed,AtkinsonK01}.
And since then, several researchers have pointed out the benefits of using multi-level modelling languages (see, e.g.,~\cite{deLara2015,MohagheghiH10,Tolvanen016,Kelly08DSMBook}).
Indeed, different authors have proposed different formalizations of multilevel modelling languages and different aspects related to multilevel structures (see, e.g.,~\cite{delara2010deep,rossini2014formalisation}).

Proposals such as MetaDepth~\cite{delara2010deep}, Melanee ~\cite{atkinson2016melanee}, AtomPM~\cite{syriani2013atompm} or Modelverse~\cite{van2014modelverse} 
propose conceptual frameworks and tools for multilevel modelling.
These approaches are based on the concept of \emph{clabject} ~\cite{atkinson1997distributed}, which provides two facets for every modelling element, so that they can be seen as classes or as objects. 
Clabjects stem from the traditional object-oriented programming, so their realization into a metamodelling framework requires a linguistic metamodel that all the levels must share.
That is, the clabject element is contained in a linguistic metamodel, together with other elements such as \emph{field} and \emph{constraint}.

MetaDepth~\cite{delara2010deep} is a deep modelling tool built on the notion of orthogonal classification and deep instantiation.
MetaDepth supports several interesting features such as model transformation reuse~\cite{deLara2015} and generic metamodelling~\cite{delara2010mixin}.
Melanee has been developed with a stronger focus on editing capabilities ~\cite{atkinson2016melanee}, as well as possible applications into the domains of executable models ~\cite{atkinson2015execution}.
AtomPM ~\cite{syriani2013atompm} is a modelling framework highly focused on the development of cloud and web tools.
Modelverse ~\cite{van2014modelverse} offers multilevel modelling functionalities by implementing the concept of clabject and building a linguistic metamodel that includes a synthetic typing relation.
In~\cite{mallet2009automated}, the same idea is applied the implementation of multilevel modelling.

All these previous approaches implement their multilevel modelling solutions by using a linguistic metamodel including the clabject element, and \emph{flattening} the ontological hierarchy as an instance of this linguistic metamodel.
That is, the whole ontological stack becomes an instance of the clabject-based modelling language.
Moreover, they require the creation of supporting tools, such as editors, constraint definition mechanisms and import/export capabilities to more widespread tools like EMF, from scratch.
With our approach, multilevel modelling is realised in a different way, which improves its flexibility and removes the need for custom-made environments and tools.
Our approach that does not require the definition of a specific linguistic metamodel, nor a flattening of the ontological levels.

\subsection{Reusability}

We have illustrated in previous sections how the use of multilevel modelling hierarchies may improve reusability. 
There are several alternative proposals to improve reusability of models and model transformations. 
In~\cite{chechik2016perspectives}, two perspectives on model transformation reuse are presented: 
programming language- and MDSE-based. 
For each perspective, the authors discuss two approaches: subtyping and mapping, and lifting and aggregating. 
In~\cite{struber2015variability} a variability-based graph transformation approach is introduced to tackle the performance problems that are introduced by systems in which a substantial number of the rules are similar to each other. 
In~\cite{sen2012reusable} an approach to reusable model transformations is presented, which is based on sub-typing an effective part of the existing source metamodel. 
That is, the metamodel of the models to be transformed is made a subtype of a pruned metamodel.
In this way, the models can be transformed by the same transformation rules which were written for the source metamodel.

To achieve genericness, in~\cite{sanchez2011generic,delara2014flexiblereuse} the rules are typed over a \emph{generic} metamodel which is called \emph{concept}.
Then, any metamodel to which there exist an embedding from the concept-metamodel (called binding) can be used by composition for the type-check during matching.
Thus, any model in a hierarchy can be typed by the concept metamodel and get the transformation rules for free.
However, it is not always straight forward to define the embedding morphism from the concept metamodel to the metamodel.
This might be because the metamodel has several structures which have the same behaviour leading to several bindings.
This is solved by introducing syntax for the definition of cardinality in which the concept metamodel can be written in a generic way, however, in the realisation of the concept, this is just syntactic sugar for the definition of multiple concept metamodels.
Moreover, finding reasonable embeddings due to structure mismatches (or heterogeneity) might be a challenge.
Adapters and concept inheritance could be seen as solution of this problem, as seen~\cite{sanchez2011generic}.

The notion of \emph{concept} from~\cite{delara2014flexiblereuse} is extended in~\cite{duran2015amalgamation} to parametric models, where the parameters have both structure and behaviour. In this case we not only have a mechanism for the reutilisation of transformations, but mechanisms for the reutilisation and composition of models with behaviour. 
The challenges in finding embeddings is however more difficult, since rules in parameter models must also be mapped to the corresponding target models.

\subsection{Multi-level model management and transformation}

Although some management operations on hierarchies have been satisfactorily formalized (see, e.g.,~\cite{rossini2014formalisation,deLara2015}), the definition of transformations on multi-level hierarchies is not yet so well handled.
Current proposals use traditional two-level transformations to deal with multi-level-hierarchy transformations.
Multi-modelling languages like the ones proposed in~\cite{JablonskiVD08} to define domain specific process modelling notations, in~\cite{ZschalerKDPR09} to express variability over DSLs and to extend them with interfaces for model reuse, or in~\cite{HerrmannsdorferH10} to declare component types with ports, use multi-level languages to express variability and reuse of multi-level languages, but it is done by emulating three meta-levels within two.

The only proposals supporting some form of deep transformations are, to some extend, Melanee~\cite{AtkinsonGK09,atkinson2012towards}, MetaDepth~\cite{delara2010deep} and DeepJava~\cite{KuhneS07}.
They follow very different approaches, but none of them provides a formalization of multi-level transformations as we do.
Here, we focus on their possibilities for multi-level hierarchy transformations.

MetaDepth~\cite{delara2010deep} is a deep modelling tool built on the notion of orthogonal classification and deep instantiation.
It is a textual tool, and provides facilities to write executable code against it.
De Lara et al. in~\cite{deLara2015} present their extension of the Epsilon~\cite{Epsilon} family of languages to support multilevel scenarios:
They define model manipulation operations, code generators working at several metalevels, and model-to-model transformations~\cite{deLara2015}.
However, all these transformations are defined at a particular level of the hierarchy, perhaps with references to indirect types.
For transformations affecting multiple levels, they propose what they call ``co-transformations'', transformations that transform a model and its metamodel at the same time.
Their are called co-transformations because they are defined as transformations that operate on respective levels, but are applied simultaneously.
Similar approaches are discussed in~\cite{HerrmannsdorferW14} in the context of co-evolution, where models and their corresponding metamodels are transformed simultaneously. 
They also provide support for refinement of transformations, when transformations require some adaptation for their application, and linguistic extensions.

Melanee~\cite{AtkinsonGK09} is a graphical multi-level modelling tool
based on EMF and GMF.
Atkinson et al. present in~\cite{atkinson2012towards} their ideas for extending the transformation language ATL to work on multi-level models.
Although they discuss the two-level to multilevel and multilevel to two-level transformation cases, the paper focuses on the multilevel to two-level case.
Moreover, in this setting, the transformation is applied at the bottommost level of the source model.
Although discussed in the paper, Melanee does not support in-place transformations.

DeepJava~\cite{KuhneS07} extends Java with the ability to perform multiple instantiations.
DeepJava programs are compiled into Java.

\subsection{In-place transformation rules for the definition and analysis of behaviour}

Two-level modelling tools such as Groove~\cite{Rensink08}, AGG~\cite{taentzer2003agg}, ATOM$^3$~\cite{LaraV02,JuanDeLara2006} and e-Motions~\cite{rivera2009graphical} use in-place model transformation to deal with the behaviour of DSLs.
The Groove and AGG tools do not support the definition of a specific graphical notation associated to the DSL at hand: rule patterns are defined with graphs. AToM$^3$ and e-Motions allow the user to specify a graphical concrete syntax for the DSLs.
From these tools, only e-Motions support the interoperation of the models defined using it with other tools. 
As MulEcore, tools like the \emph{e-Motions} graphical framework, Tiger EMF~\cite{GraphicalDefinitionInPlaceRules} or MOMENT2~\cite{Boronat-Meseguer:2009}
were developed for the Eclipse platform, and enable their integration with other modelling tools already defined for Eclipse, such as EMF, GMF, ATL, TCS, etc., because they all share the same representation/notation of models and metamodels. 
This allows users to benefit from all of these tools to achieve a complete MDSE process.
Tools like Groove, AGG and ATOM$^3$ are developed as stand-alone tools, i.e., their connection with other modelling tools requires a high effort, because they do not share a common model notation.

Systems like e-Motions and Groove focus on the definition of timed systems, and on the verification of the systems specified, providing facilities for model-checking or statistical model-checking inside their environments~\cite{ghamarian2012groove,duran2016statistical}. 
See~\cite{Rensink08,ghamarian2012groove} for a discussion on Groove, or~\cite{SimulationMaude} for the use of different formal verification techniques on systems specified using in-place model transformations.

\section{Conclusions and future work}
\label{sec:conclusions-and-future-work}

This paper presents an approach for multilevel modelling based on graphs and Category Theory constructions, such as pushouts and pullbacks in the corresponding category.
The approach tackles both multilevel metamodelling --- the definition of hierarchies --- and multilevel model transformation --- the definition of transformation rules --- with focus on specifying and reusing behaviour in the same domain and among similar ones.
The main focus of the work presented here are in-place, multilevel model transformations, for which we coin the concept of Multilevel Coupled Model Transformations, MCMTs. This framework is presented conceptually (in Section 3), theoretically (in Section 4) and in practice (in Section 5).

We have argued how MCMTs can achieve greater expressive power than traditional, two-level MTs, which allows to reuse behaviour for different elements in the same model and, more importantly, for different models with shared commonalities.
This reusability reduces the amount of transformation rules that need to be defined, as detailed in Section~\ref{subsec:why-using-mcmt}.
In this regard, the 4 rules presented in Section~\ref{subsec:mcmt-pls} (\textit{CreatePart}, \textit{SendPartOut}, \textit{Assemble} and \textit{Transferpart}) generate, after the proliferation process for the typing chain of \elementname{hammer\_config}, 21 two-level rules, which would have needed to be manually specified otherwise.
This number of rules would further increase if we consider defining behaviour for the typing chain of \elementname{stool\_plant} (the right branch of the PLS hierarchy), while our 4 original MCMTs would still suffice to represent the same information.

We have presented a formalization of MLM hierarchies, on which we have introduced our MCMTs. 
Since every branch in our MLM hierarchies is represented by a graph chain, the application of our model transformation rules are formalized by appropriate pushout and pullback complement constructions in the category \cat{Chain} of graph chains and graph chain morphisms. 
Specifically, our constructions for graph chains build on the well-established constructions of pushout and pullback complement in the lowest level graphs of the hierarchy in the category \cat{Graph}, and indeed get the multilevel typing of the constructed graphs.

When it comes to the usage of potency, we have also redefined its semantics, compared to other approaches like~\cite{rossini2014formalisation}.
The range-like potency specification allows us to constrain a type more specifically than the single-valued one, with the extra advantage that a single notation covers both notations defined in those approaches (jump-like and continuous).
Additionally, our version of potency allows to specify values like \typename{}{2-3}, which are not possible to replicate with single-valued ones.

About the next steps in our work, we plan to increase the tool support to both hierarchies and MCMTs specification, as well as test their feasibility in new scenarios.
In addition, we plan to use, or adapt, strategy languages like~\cite{krause2014implementing,fernandez2011strategy} in order to coordinate the execution of MCMTs, allowing for even richer semantics for the execution of behavioural DSLs~\cite{macias2017chains}.

\bibliographystyle{elsarticle-num}
\bibliography{bibliography}

\end{document}